%
%
%

\input harvmac

\def\lesssim{\mathrel{\mathpalette\fun <}}

\def\fun#1#2{\lower3.6pt\vbox{\baselineskip0pt\lineskip.9pt
  \ialign{$\mathsurround=0pt#1\hfil##\hfil$\crcr#2\crcr\sim\crcr}}}
\relax

\noblackbox

\lref\HellingKZ{ R.~Helling, ``Beyond eikonal scattering in
M(atrix)-theory,'' hep-th/0009134.
}

\lref\EastherQS{ R.~Easther, J.~Khoury and K.~Schalm, ``Tuning
Locked Inflation: Supergravity versus Phenomenology,''
hep-th/0402218.
}

\lref\BrandleFA{
M.~Brandle and A.~Lukas,
``Flop transitions in M-theory cosmology,''
Phys.\ Rev.\ D {\bf 68} (2003) 24030, hep-th/0212263.
}

\lref\BanksDP{
T.~Banks, M.~Berkooz, S.~H.~Shenker, G.~W.~Moore and P.~J.~Steinhardt,
Phys.\ Rev.\ D {\bf 52} (1995) 3548,
hep-th/9503114.
}

\lref\LindeXX{ A.~D.~Linde, D.~A.~Linde and A.~Mezhlumian, ``From
the Big Bang theory to the theory of a stationary universe,''
Phys.\ Rev.\ D {\bf 49} (1994) 1783, gr-qc/9306035\semi
J.~Garcia-Bellido, A.~D.~Linde and D.~A.~Linde, ``Fluctuations of
the gravitational constant in the inflationary Brans-Dicke
cosmology,'' Phys.\ Rev.\ D {\bf 50} (1994) 730, astro-ph/9312039\semi J.~Garriga and A.~Vilenkin, ``A
prescription for probabilities in eternal inflation,'' Phys.\
Rev.\ D {\bf 64} (2001) 23507, gr-qc/0102090\semi A.~Linde, ``Inflation, quantum cosmology and the anthropic
principle,'' in {\it{Science and Ultimate Reality: From Quantum to Cosmos}},'' J.D. Barrow, P.C.W. Davies, and C.L. Harper, eds.,
Cambridge University Press, Cambridge (2003), hep-th/0211048\semi J.~Garriga, A.~Linde and
A.~Vilenkin, ``Dark energy equation of state and anthropic
selection,'' hep-th/0310034.}

\lref\KofmanFI{ L.~Kofman, A.~D.~Linde and A.~A.~Starobinsky,
``Non-Thermal Phase Transitions After Inflation,'' Phys.\ Rev.\
Lett.\  {\bf 76} (1996) 1011, hep-th/9510119\semi  G.~N.~Felder,
L.~Kofman, A.~D.~Linde and I.~Tkachev, ``Inflation after
preheating,'' JHEP {\bf 0008} (2000) 10, hep-ph/0004024.
}

\lref\StromingerPC{ A.~Strominger, ``Open string creation by
S-branes,'' hep-th/0209090.
}

\lref\MaloneyCK{ A.~Maloney, A.~Strominger and X.~Yin, ``S-brane
thermodynamics,'' JHEP {\bf 0310} (2003) 048, hep-th/0302146.
}

\lref\BachasKX{ C.~Bachas, ``D-brane dynamics,'' Phys.\ Lett. {\bf{B}}374
(1996) 37, hep-th/9511043.
}

\lref\BoussoXA{ R.~Bousso and J.~Polchinski,
 ``Quantization of four-form fluxes and dynamical neutralization of the
cosmological constant,'' JHEP {\bf 0006} (2000) 006, hep-th/0004134.
}

\lref\KhlebnikovMC{ S.~Y.~Khlebnikov and I.~I.~Tkachev,
``Classical decay of inflaton,'' Phys.\ Rev.\ Lett.\ {\bf 77}, 219
(1996), hep-ph/9603378.
}

\lref\FelderHQ{ G.~N.~Felder and I.~Tkachev, ``LATTICEEASY:  A
program for lattice simulations of scalar fields in an expanding
universe,'' hep-ph/0011159.
}

\lref\douglas{M. R. Douglas, ``The Statistics of String/M Theory
Vacua,'' JHEP {\bf{0305}} (2003) 046, hep-th/0303194; S. Ashok and
M. R. Douglas, ``Counting Flux Vacua,'' hep-th/0307049.}

\lref\rs{ L. Randall and R. Sundrum, ``A Large Mass Hierarchy from
a Small Extra Dimension,'' Phys.Rev.Lett.{\bf{83}} (1999) 3370,
hep-ph/9905221.
}

\lref\FelderJK{ G.~N.~Felder, A.~V.~Frolov, L.~Kofman and
A.~D.~Linde, ``Cosmology with negative potentials,'' Phys.\ Rev.\
{\bf{D}}  66 (2002) 023507, hep-th/0202017.
}

\lref\MotlCY{T. Banks, W. Fischler, and L. Motl, ``Dualities
versus  Singularities,''  JHEP {\bf{9901}} (1999) 019,
hep-th/9811194 \semi L.~Motl and T.~Banks, ``On the hyperbolic
structure of moduli spaces with 16 SUSYs,'' JHEP {\bf 9905} (1999) 015,
hep-th/9904008.
}

\lref\DouglasYP{ M.~R.~Douglas, D.~Kabat, P.~Pouliot and
S.~H.~Shenker,  ``D-branes and short distances in string theory,''
Nucl.\ Phys.\ {\bf{B}}485 (1997) 85, hep-th/9608024.
}

\lref\achar{
B.~S.~Acharya, ``A moduli fixing mechanism in M theory,'' hep-th/0212294.
}

\lref\KKLT{
S.~Kachru, R.~Kallosh, A.~Linde and S.~P.~Trivedi, ``de Sitter
Vacua in String Theory,'' Phys.\ Rev.\ D {\bf 68} (2003) 046005, hep-th/0301240.
}

\lref\MSS{
A.~Maloney, E.~Silverstein and A.~Strominger, ``De Sitter space
in noncritical string theory,'' hep-th/0205316;
E.~Silverstein, ``(A)dS backgrounds from asymmetric
orientifolds,'' hep-th/0106209.
}

\lref\GukovCY{ S.~Gukov, S.~Kachru, X.~Liu and L.~McAllister,
``Heterotic Moduli Stabilization with Fractional Chern-Simons
Invariants,'' hep-th/0310159.
}

\lref\kls{L.~Kofman, A.~D.~Linde and A.~A.~Starobinsky, ``Towards
the theory of reheating after inflation,'' Phys.\ Rev.\ D {\bf
56}, 3258 (1997), hep-ph/9704452.
}

\lref\gs{G. Dvali and S. Kachru, ``New Old Inflation,'' hep-th/0309095.
}

\lref\SilversteinHF{ E.~Silverstein and D.~Tong,  ``Scalar speed
limits and cosmology: Acceleration from D-cceleration,''
hep-th/0310221; M. Alishahiha, E. Silverstein, and D. Tong, in
progress.
}

\lref\jarv{ L. Jarv, T. Mohaupt, and F. Saueressig,  ``Effective
Supergravity Actions for Flop Transitions,'' JHEP {\bf{0312}}
(2003) 047, hep-th/0310173 \semi L. Jarv, T. Mohaupt, and F.
Saueressig, ``M-theory Cosmologies from Singular Calabi-Yau
Compactifications,'' hep-th/0310174.}

\lref\bd{N. D. Birrell and P.C.W. Davies,  {\it Quantum Fields in
Curved Space}, Cambridge University Press, Cambridge, England
(1982).}

\lref\spacebranes{string, D creation in tachyon decay}

\lref\horne{J. Horne and G. Moore, ``Chaotic Coupling Constants,''
 Nucl.Phys. {\bf{B}}432 (1994) 109, hep-th/9403058.}

\lref\dine{M. Dine, ``Towards a Solution of the Moduli  Problems
of String Cosmology,'' Phys.Lett. {\bf{B}}482 (2000) 213,
hep-th/0002047 \semi M. Dine, Y. Nir, and Y. Shadmi, ``Enhanced
Symmetries and the Ground State of String Theory,'' Phys.Lett.
{\bf{B}}438 (1998) 61, hep-th/9806124.}

\lref\hyb{A. Linde, ``Hybrid Inflation,'' Phys.Rev. {\bf{D}}49
(1994)  748, astro-ph/9307002.}

\lref\sw{N. Seiberg and E. Witten, ``Electric-Magnetic Duality,
Monopole Condensation, and Confinement in N=2 Supersymmetric
Yang-Mills Theory,'' Nucl.Phys. {\bf{B}}426 (1994) 19,
hep-th/9407087 \semi N. Seiberg and E. Witten, `` Monopoles,
Duality and Chiral Symmetry Breaking in N=2 Supersymmetric QCD,''
Nucl.Phys.{\bf{B}}431 (1994) 484, hep-th/9408099 \semi K.
Intriligator and N. Seiberg, ``Lectures on Supersymmetric Gauge
Theories and Electric-Magnetic Duality,'' Nucl.Phys.Proc.Suppl.
45BC (1996) 1, hep-th/9509066.}

\lref\chung{D.J.H. Chung, E.W. Kolb, A. Riotto, and I.I. Tkachev,
``Probing Planckian Physics: Resonant Production of Particles
During Inflation and Features in the Primordial Power Spectrum,''
Phys.Rev. {\bf{D}}62 (2000) 043508, hep-ph/9910437.}

\lref\flux{some papers on moduli stabilization from fluxes}

\lref\mavan{R. Fardon, A.E. Nelson, and N. Weiner, ``Dark Energy
from Mass Varying Neutrinos,'' astro-ph/0309800.}

\lref\gubser{S.S. Gubser, ``String Production at the Level of
Effective Field Theory,'' hep-th/0305099; ``String Creation and
Cosmology,'' hep-th/0312321\semi J.J. Friess, S.S. Gubser, and I. Mitra,
``String Creation in Cosmologies with a Varying Dilaton,'' hep-th/0402156.}

\lref\noinflation{G. Felder, L. Kofman, and A. Linde, ``Inflation
and Preheating in NO Models,'' Phys.Rev. {\bf{D}}60 (1999) 103505,
hep-ph/9903350.}

\lref\conifold{A. Strominger, ``Massless Black Holes and Conifolds
in String Theory,'' Nucl.Phys. {\bf{B}}451 (1995) 96,
hep-th/9504090.}

\lref\ADE{E. Witten, ``String Theory Dynamics In Various
Dimensions,'' Nucl.Phys. {\bf{B}}443 (1995) 85, hep-th/9503124
\semi S. Katz, D. R. Morrison, and M. R. Plesser, ``Enhanced Gauge
Symmetry in Type II String Theory,'' Nucl.Phys. {\bf{B}}477
(1996) 105, hep-th/9601108 \semi M. Bershadsky, K. Intriligator,
S. Kachru, D.R. Morrison, V. Sadov, and C. Vafa,  ``Geometric
Singularities and Enhanced Gauge Symmetries,'' Nucl.Phys.
{\bf{B}}481 (1996) 215, hep-th/9605200.}

\lref\smallinstanton{E. Witten, ``Small Instantons in String
Theory,''  Nucl.Phys. {\bf{B}}443 (1995) 85, hep-th/9511030.}

\lref\WittenIM{ E.~Witten, ``Bound states of strings and
p-branes,''  Nucl.\ Phys.{\bf{B}}460 (1996) 335,
hep-th/9510135.
}

\lref\instant{G.~N.~Felder, L.~Kofman and A.~D.~Linde, ``Instant
preheating,'' Phys.\ Rev.\ D {\bf 59}, 123523 (1999), hep-ph/9812289.}

\lref\FelderHJ{ G.~N.~Felder, J.~Garcia-Bellido, P.~B.~Greene,
L.~Kofman, A.~D.~Linde and I.~Tkachev, ``Dynamics of symmetry
breaking and tachyonic preheating,'' Phys.\ Rev.\ Lett.\  {\bf
87} (2001) 011601, hep-ph/0012142.
}

\lref\FelderHR{ G.~N.~Felder and L.~Kofman, ``The development of
equilibrium after preheating,'' Phys.\ Rev.\ D {\bf 63}, 103503
(2001), hep-ph/0011160.
}


\lref\berera{ A.~Berera and T.~W.~Kephart, ``Ubiquitous  inflaton
in string-inspired models,'' Phys.\ Rev.\ Lett.\  {\bf 83} (199) 1084,
hep-ph/9904410.
}

\lref\SUSY{ A.~D.~Linde, ``Inflation Can Break Symmetry In SUSY,''
Phys.\ Lett.\ B {\bf 131} (1983) 330.
}

\lref\LindeDQ{ A.~D.~Linde, ``Inflation And Quantum Cosmology,''
Print-86-0888 (June 1986), in {\it 300 Years of Gravitation},
S.W. Hawking and W. Israel, eds., Cambridge University Press,
Cambridge (1987).
}

\input epsf
\noblackbox
\newcount\figno
\figno=0
\def\fig#1#2#3{
\par\begingroup\parindent=0pt\leftskip=1cm\rightskip=1cm\parindent=0pt
\baselineskip=11pt \global\advance\figno by 1 \midinsert
\epsfxsize=#3  \centerline{\epsfbox{#2}} \vskip 12pt {\bf Fig.\
\the\figno: } #1\par
\endinsert\endgroup\par
}
\def\figlabel#1{\xdef#1{\the\figno}}

\def\p{\partial}
\def\vphi{{ \phi}}
\def\ra{\rangle}
\def\la{\langle}

\Title{\vbox{\baselineskip12pt\hbox{hep-th/0403001}
\hbox{CITA-04-3}\hbox{SLAC-PUB-10343}\hbox{SU-ITP-04/05}
}}{\vbox{\vskip -1.5cm\centerline {Beauty is Attractive:} \vskip
0.5cm \centerline{ Moduli Trapping at Enhanced Symmetry Points }}}

\vskip -0.3cm

\centerline{Lev Kofman,$^a$ Andrei Linde,$^b$ Xiao Liu,$^{b,c}$}
\centerline{ Alexander Maloney,$^{b,c}$ Liam McAllister,$^{b,c}$
and Eva Silverstein$^{b,c}$}

\medskip
\centerline{$^a$ \it CITA, University of Toronto, Toronto, ON M5S
3H8, Canada} \centerline{$^b$ \it Department of Physics, Stanford
University, Stanford, CA 94305, USA} \centerline{$^c$ \it SLAC,
Stanford University, Stanford, CA 94309, USA}
\vskip .3in  We study quantum effects on moduli dynamics arising
from the production of particles which are light at special points
in moduli space. The resulting forces trap the moduli at these points, which often
exhibit enhanced symmetry.  Moduli trapping
occurs in time-dependent quantum field theory, as well as in
systems of moving D-branes, where it leads the branes to combine
into stacks. Trapping also occurs in an expanding universe, though
the range over which the moduli can roll is limited by Hubble
friction. We observe that a scalar field trapped on a steep
potential can induce a stage of acceleration of the universe,
which we call trapped inflation. Moduli trapping ameliorates the
cosmological moduli problem and may affect vacuum selection. In
particular, rolling moduli are most powerfully attracted to the
points with the largest number of light particles, which are often
the points of greatest symmetry. Given suitable assumptions about
the dynamics of the very early universe, this effect might help to
explain why among the plethora of possible vacuum states of string
theory, we appear to live in one with a large number of light
particles and (spontaneously broken) symmetries. In other words,
some of the surprising properties of our world might arise
not through pure chance or miraculous cancellations, but through
a natural selection mechanism during dynamical evolution.

\smallskip
\Date{} \listtoc \writetoc


\newsec{Introduction}
\subsec{Moduli Trapping Near Enhanced Symmetry Points}

Supersymmetric string and field theories typically contain a
number of light scalar fields, or moduli, which describe
low-energy deformations of the system.  If the kinetic energy of
these fields is large compared to their potential energy then the
classical dynamics of the moduli is described by geodesic motion
on moduli space.

At certain special points (or subspaces) of moduli space, new
degrees  of freedom become light and can affect the dynamics of
moduli in a significant way
\refs{\kls,\instant,\noinflation,\chung,\SilversteinHF}. These
extra species often contribute to an enhanced symmetry at the
special point. We will refer to any points where new species
become light as ESPs, which stands for extra species points, and
also, when applicable, for enhanced symmetry points.

A canonical example is a system of two parallel D-branes. When the
branes coincide, the two individual $U(1)$ gauge symmetries are
enhanced to a $U(2)$ symmetry, as the strings that stretch between
the branes become massless \WittenIM.  Similar points with new
light species arise in many contexts; examples include the
Seiberg-Witten massless monopole and dyon points in ${\cal N}=2$
supersymmetric field theories \sw, the conifold point \conifold\
and ADE singularities in Calabi-Yau compactification \ADE, the
self-dual radius of string compactifications on a torus, small
instantons in heterotic string theory \smallinstanton, and many
other configurations with less symmetry.

Classically, there is no sense in which these ESPs are dynamically
preferred over other metastable vacuum states of the system. We
will argue that this changes once quantum effects are included. In
particular, quantum particle production of the light fields alters
the dynamics in such a way as to drive the moduli towards the ESPs
and trap them there.

The basic mechanism of this trapping effect is quite simple.
Consider a modulus $\phi$ moving through moduli space near an ESP
associated to a new light field $\chi$.  For example, $\phi$ could
be the separation between a pair of parallel D-branes, and $\chi$
a string stretching between the two branes  -- in this case the
ESP $\phi=0$ is the point where the branes coincide and $\chi$
becomes massless. As $\phi$ rolls through moduli space, the mass
of $\chi$ changes; $\chi$ gets lighter as $\phi$ moves closer to
the ESP and heavier as $\phi$ moves farther away. This changing
mass leads to quantum production of $\chi$ particles; as $\phi$
moves past the ESP some of its kinetic energy will be dumped into
$\chi$ particles. As $\phi$ rolls away from the ESP, more and more
of its energy will be drained into the $\chi$ sector as the $\chi$
mass increases, until eventually $\phi$ stops rolling. At this
point the moduli space approximation for $\phi$ has broken down,
and all of the original kinetic energy contained in the coherent
motion of $\phi$ has been transferred into $\chi$ particles, and
ultimately into all of the fields interacting with $\chi$
(including decoherent quanta $\delta \phi$). As we will see in
detail, the $\chi$ excitations generate a classical potential for
$\phi$ which drives the modulus back toward the ESP and traps it
there.\foot{There are also corrections to the effective action for
$\phi$ from loops of $\chi$ particles, including both kinetic
corrections and a Coleman-Weinberg effective potential. Both
effects will be subdominant in the weakly-coupled, supersymmetric,
kinetic-energy dominated regimes we will consider.}

In the example of the pair of moving D-branes, the consequences of
this are simple: two parallel branes that are sent towards each
other will collide and remain bound together.  The original
kinetic energy of the moving branes will be transferred into open
string excitations on the branes and eventually into closed
string radiation in the bulk.

In \S2 we will describe the general trapping mechanism and study
its  range of applicability using a few simple estimates. In \S3
we will write down the equations of motion governing trapping in
more detail, and describe the numerical and analytic solutions of
these equations in a variety of cases.

It is important to recognize that this trapping effect is in no
way special to string theory.  Flat space quantum field theory
with a moduli space for $\phi$ and an ESP is an ideal setting for
the trapping  effect, and it is in this setting that we will
perform the analysis of \S2 and \S3. In \S4 we will generalize
this to incorporate the effects of cosmological expansion, and in
\S5\ we will discuss the possibility of significant effects from
string theory.  Having established the moduli trapping effect in a
variety of contexts, we will then study its applications to
problems in cosmology.

The most immediate application is to the problem of vacuum
selection. As we will see in \S6, the trapping effect can provide
a dynamical vacuum selection principle, reducing the problem to
that of selecting one point within the class of ESPs.  This
represents significant progress, since the vast majority of
metastable vacua are not ESPs. Trapping at ESPs may also help
solve the cosmological moduli problem, as we will see in \S7. In
particular, trapping strengthens the proposal of \dine\ by
providing a dynamical mechanism which explains why moduli sit at
points of enhanced symmetry.

Finally, as we will explain in \S8, the trapping of a scalar field
with  a potential can lead to a period of accelerated expansion,
in a manner reminiscent of thermal inflation \berera. This effect,
which we will call trapped inflation, can occur in a steeper
potential than normally admits such behavior.

From a more general perspective, moduli trapping gives us insight
into the celebrated question of why the world is so symmetric. The
initial puzzle is that although highly symmetric theories are
aesthetically appealing and theoretically tractable, they are also
very special and hence, in an appropriate sense, rare. One expects
that in a typical string theory vacuum, most symmetries will be
strongly broken and most particles will have masses of order the
string or Planck mass, just as in a typical vacuum one expects a
large cosmological constant. Vacua with enhanced symmetry or light
particles should comprise a minuscule subset of the space of all
vacua.

Nevertheless, we observe traces of many symmetries in the
properties of elementary particles, as spontaneously broken global
and gauge invariances.  Moreover, all known particles are
hierarchically light compared to the Planck mass. Given the
expectation that a typical vacuum contains very few approximate
symmetries and very few light particles, it is puzzling that we
see such symmetries and such particles in our world.

For questions of this nature, moduli trapping may have
considerable  explanatory power. Specifically, the force pulling
moduli toward a point of enhanced symmetry is proportional to the
number of particles which become massless at this point, which is
often associated with a high degree of symmetry. This means that
the most attractive ESPs are typically
the ones with the largest symmetry, and rolling moduli are most
likely to be trapped at highly symmetric points, where many
particles become massless or nearly massless. Moreover, the
process of trapping can proceed sequentially: a modulus moving in
a multi-dimensional moduli space can experience a sequence of
trapping events, each of which increases the symmetry.  These
effects suggest that the symmetry and beauty we see in our world
may have, at least in part, a simple dynamical explanation: beauty
is attractive. We will discuss this possibility in \S6.

\subsec{Relation to Other Works}

Similar effects have been described in the literature. There has
been much work on multi-scalar quantum field theory in the context
of inflation, especially concerning preheating  in interacting
scalar field theories. Some of our results will be based on the
theory of particle production and preheating developed in the
series of papers \refs{\kls,\instant,\noinflation}, which explores
many of the basic phenomena in scalar theories of the sort we will
consider. Likewise, Chung et al. \chung\ have explored the effects
of particle production on the inflaton trajectory and on the
spectrum of density perturbations. Although we will derive what we
need here in a self-contained way, many of the technical results
in this paper overlap with those works, as well as with standard
results on particle production in time-dependent systems as
summarized in e.g. \bd.  Although we will not study the case in
which $\chi$ goes tachyonic for some range of $\phi$, our results
may nevertheless have application to models of hybrid inflation
\refs{\hyb,\FelderHJ}, including models based on rapidly-oscillating
interacting scalars \refs{\KofmanFI,\gs,\EastherQS}.

In strong 't Hooft coupling regions of moduli spaces which are
accessible through the AdS/CFT correspondence, virtual effects
from the large numbers of light species dramatically slow down the
motion of $\phi$ as it approaches an ESP, with the result that the
modulus gets trapped there \SilversteinHF. This also provides a
mechanism for slow roll inflation without very flat potentials. In
the present work, which applies at weak 't Hooft coupling, it is
quantum production of {\it{on-shell}} light particles which leads
to trapping on
moduli space.

Other works in the context of string theory have explored the
localization  of moduli at ESPs.  The authors of \refs{\jarv,\BrandleFA} 
studied the evolution of a supersymmetric version of the $\phi-\chi$ 
system arising near a flop transition using an effective 
supergravity action. They showed that, given nonvanishing initial
vevs for both $\phi$ and $\chi$, the fields will settle at the ESP
even if one formally turns off particle production effects.  Our
proposal, by contrast, is to take into account on-shell quantum
effects which dynamically generate a nonzero $\la \chi^2 \ra$. In
works such as \MotlCY\ attention was focused on the boundaries of
moduli space, while here we focus on ESPs in the interior of
moduli space. In \DouglasYP, production of light strings was
studied in the context of D0-brane quantum mechanics; as we
explain in \S2.3, this has some similarities, but important
differences, with our case of space-filling branes.  Scattering of
Dp-branes was also studied in \BachasKX.

Dine has suggested that enhanced symmetry points may provide a
solution to the moduli problem, as moduli which begin at an
enhanced symmetry minimum of the quantum effective potential can
consistently remain there both during and immediately after
inflation \dine.  One would still like to explain why the moduli
began at such a point. As we discuss in \S7, our trapping
mechanism provides a natural explanation for this initial
configuration.

Horne and Moore \horne\ have argued that the classical motion on certain
moduli spaces is ergodic, provided that the potential energy is
negligible.  This means that all configurations are sampled given a
sufficiently long time, and in particular a given modulus will eventually
approach an ESP.  We will argue that quantum corrections to the classical
trajectory are significant, and indeed lead to trapping, whenever the
classical trajectory comes close to an ESP.  Combining these two
observations, we expect that in the full, quantum-corrected system the
moduli are stuck near an ESP at late times.  This means that the
quantum-corrected evolution is not fully ergodic: the dynamics of \horne\
(see also \BanksDP) implies that the modulus will eventually approach an
ESP, at which point quantum effects will trap it there, preventing the
system from sampling any further regions of moduli space.

\newsec{Moduli Trapping: Basic Mechanism}

We will now describe the mechanism of moduli trapping in more
detail. Our discussion in this section will be based on simple
estimates of particle production and the consequent backreaction,
generalizing the results of  \refs{\kls,\instant,\noinflation}
to the case of a complex field. A more complete analysis, along with
numerical results, will be presented in \S3.

We will consider the specific model
\eqn\lis{ {\cal L} = {1\over2} \p_\mu{\vphi} \p^\mu\bar\vphi +
{1\over 2}\p_\mu \chi \p^\mu \chi  - {g^2\over 2} |\vphi|^2
\chi^2 }
where a complex modulus $\vphi=\phi_1+i\phi_2$ interacts with a
real scalar field $\chi$. We are restricting ourselves to the case
of a flat moduli space which has a single ESP at $\vphi=0$, where
$\chi$ becomes massless, and a particularly simple form for the
$\chi$ interaction. This simple case illustrates the basic physics
and can be generalized as necessary, for example to include
supersymmetry.

We will consider the case where $\vphi$ approaches the origin with
some impact parameter $\mu$, following a classical trajectory of
the form \eqn\phiis{
\vphi (t) = i \mu +{v} t .} Classically, if $\chi$ vanishes then
\phiis\ is an exact solution to the equations of motion, and the
presence of the ESP will not affect the motion of $\phi$.

Quantum effects will alter this picture considerably, because the
trajectory \phiis\ will lead to the production of $\chi$
particles, as we discuss in \S2.1.  The backreaction of these
particles on the motion of $\phi$ will then lead to trapping, as
we will see in \S2.2.  In \S2.3 we will illustrate this effect
with the example of colliding D-branes.

\subsec{Quantum Production of $\chi$ Particles}

Let us first study the creation of $\chi$ particles without
considering how they may backreact to alter the motion of $\vphi$.
In this approximation we may substitute \phiis\ into the action
\lis\ to get a free quantum field theory for $\chi$ with a
time-varying mass \eqn\mwis{ m^2_{\chi}(t) = g^2 |\vphi(t)|^2 .}
This time dependence leads to particle production.

Consider a mode of the  $\chi$ field with spatial momentum $k$,
whose frequency \eqn\freq{ \omega(t) = \sqrt{k^2 +
g^2|\vphi(t)|^2} } varies in time.  This mode becomes excited when
the non-adiabaticity parameter $\dot\omega/\omega^2$
becomes at least of order one.
This parameter vanishes as $t\to\pm\infty$, indicating that
particle creation takes place only while $\phi$ is near the ESP.
It is straightforward to see that, for the trajectory \phiis,
$\dot\omega/\omega^2$ can be large only in the small interval
$|\phi| \lesssim \Delta \phi$ near the ESP, where
\eqn\deltaphi{ \Delta\phi = \sqrt{v\over g},}
and only for momenta
\eqn\range{ {k^2 + g^2 \mu^2 \over gv} \lesssim 1 .}
When the quantity on the left hand side is small, particle
creation effects are very strong. They are strongest if the
modulus passes sufficiently close to the ESP, i.e. if
\eqn\mubound{ \mu \lesssim \sqrt{v/g}. } In this case
$\chi$ modes whose momenta $k$ fall in the range \range\ will be
excited.\foot{This may be checked as follows. We have argued that
unsuppressed particle production occurs only when the modulus is
sufficiently close to the ESP, $|\phi|\lesssim \sqrt{v/g}$.
 The modulus remains within this window for a time
\eqn\timeof{\Delta t \sim  {\sqrt{v / g} \over v} \sim (g
v)^{-1/2} . }
The uncertainty principle implies in this case that the created
particles will have typical energy $E \sim (\Delta t)^{-1}$ and
thus momenta $k \sim (g v - g^2 \mu^2 )^{1/2}$. This agrees with
the estimate \range.} Qualitatively, we expect that the occupation
numbers $n_k$ of such modes will vary from zero (no real
particles) for modes with vanishing non-adiabaticity to of order
unity for modes with very large non-adiabaticity.  The full
computation of $n_k$ given in Appendix A yields \eqn\nkis{ n_k =
\exp\left(-\pi{k^2 + g^2 \mu^2 \over gv}\right),} which agrees
with this qualitative expectation.  Note that even when \mubound\
is not satisfied, there is generically a nonvanishing, though
exponentially suppressed, number density of created particles;
even in this case we will find a nontrivial trapping effect.

Before discussing the backreaction due to the production of $\chi$
particles, it is crucial to control other effects from the $\chi$
field. In particular, there is another important quantum effect
which arises in motion toward the origin: loops of light $\chi$
particles give corrections to the effective action. These include
both kinetic corrections and the Coleman-Weinberg potential
energy. The latter we will subtract by hand, as we will explain in \S3.1.
This gives a good approximation to the dynamics in any situation where kinetic
energy dominates.

The kinetic corrections are organized in an expansion in
$v^2/\phi^4$ \SilversteinHF. The parameters controlling both
remaining effects -- the nonadiabaticity controlling particle
production and the kinetic factor $v^2/\phi^4$ controlling light
virtual $\chi$ particles -- diverge as we approach the origin.
However, at weak coupling, the nonadiabaticity parameter is
parametrically enhanced relative to the kinetic corrections, i.e.
$v^2/{g^2\phi^4} \gg v^2/\phi^4$, so we can sensibly focus on the
effects of particle production.  More specifically, we can ensure
that the kinetic corrections are insignificant by including a
sufficiently large impact parameter $\mu$.

We will also analyze the case of small $\mu$, including $\mu=0$.
This relies on the plausible assumption that the effects of the
kinetic corrections remain subdominant as we approach very close
to the origin, and that in particular in our weak coupling case
they do not by themselves stop $\phi$ from progressing through the
origin. It would be interesting to develop theoretical tools to
analyze this issue more directly and check this hypothesis.

\subsec{Backreaction on the Motion of $\phi$}

One might expect {\it{a priori}} that any description of the motion
of $\phi$ which fully incorporates backreaction from particle production
would be immensely complicated. Fortunately, this turns out not to be the case,
and a simple description is possible. The key simplification is that creation of
$\chi$ particles happens primarily in a small vicinity of the ESP
$\phi = 0$, so one can treat this as an instant event of particle
production. These particles induce a very simple linear, confining
potential acting on $\phi$,~ $V \sim |\phi|$. The motion
of $\phi$ in this potential between successive events of
particle production can be described rather simply.

Let us now explore this in more detail.  We have seen that as
$\phi$ moves in moduli space, some of its energy will be
transferred into excitations of $\chi$. This leads to a quantum
vacuum expectation value $\langle \chi^2 \rangle \ne 0$. As $\phi$
rolls away from the ESP, the mass of the created $\chi$ particles
increases, further increasing the energy contained in the $\chi$
sector.  At this point the backreaction of the $\chi$ field on the
dynamics of $\phi$ becomes important, and the moduli space
approximation breaks down.

We will concentrate on the backreaction of the created particles
on the motion of the field $\phi$ far away from the small region
of  non-adiabaticity, i.e. for $\phi \gg \Delta \phi \sim
\sqrt{v/g}$. At this stage the typical momenta are such that the $\chi$
particles are nonrelativistic, $k \lesssim \sqrt{gv} \ll
g|\phi|$. Therefore the total energy density of the gas of
$\chi$ particles is easily seen to be
\eqn\endens{\rho_{{}_{\chi}}(\phi) = \int {d^3k\over (2\pi)^3} n_k
\sqrt{k^2+ g^2|\phi(t)|^2}\approx  g|\phi(t)| n_{{}_{\chi}},  }
where $n_\chi$ is the number density of $\chi$ particles,
\eqn\numdens{ n_{{}_{\chi}} = \int {d^3k\over (2\pi)^3}  n_k =
{(gv)^{3/2}\over{(2\pi)^3}} e^{-\pi g \mu^2/v} }
As $\phi$ continues to move away from the ESP $\phi = 0$, the
number density of $\chi$ particles remains constant, as particles are
produced only in the vicinity of $\phi = 0$.  However, the
energy density of the $\chi$ particles grows as $g|\phi(t)|
n_{{}_{\chi}}$. This leads to an attractive force of magnitude $g
n_{{}_{\chi}}$, which always points towards the ESP $\phi =
0$.

This force of attraction slows down the motion of $\phi$, and eventually
turns $\phi$ back toward the ESP.  This reversal occurs in the vicinity of
the point $\phi_*$ at which the initial kinetic energy density ${1\over
2}\dot\phi^2\equiv {1\over 2} v^2$ matches the energy density
$\rho_{{}_{\chi}}$ contained in $\chi$ particles. We find
\eqn\matchen{\phi_*={{4 \pi^3}\over{g^{5/2 }}}{v^{1/2} }e^{\pi g
\mu^2/v}.}
Observe that for $g \ll 1$ the trapping length on the first
pass is always much greater than the impact parameter $\mu$, which means
that the motion of the moduli after the first impact is effectively
one-dimensional.


After changing direction at $\phi_{*}$, $\phi$ falls back toward
the  origin. On this second pass by the ESP, more $\chi$ particles
are produced, leading to a stronger attractive force. This process
repeats itself, leading ultimately to a trapped orbit of $\phi$
about the ESP, in a trajectory determined by the effective
potential and consistent with angular momentum conservation on
moduli space.

We conclude that, in this simplified setup, a scalar field which
rolls past an ESP will oscillate about the ESP with an initial
amplitude given by \matchen.

In fact, in many cases the amplitude of these oscillations will
rapidly decrease due to the effect of parametric resonance,
similar to the effects studied in the theory of preheating \kls,
and the field $\phi$ will fall swiftly towards the ESP.  This
important result will be described in more detail in \S3.3.

So far we have not incorporated the effects of scattering and
decay of the $\chi$ particles. These could weaken the trapping
potential \endens\ by reducing the number of $\chi$ particles.
Specifically, the energy density $\rho_{{}_{\chi}}$ contained in a
fixed number of $\chi$ particles \endens\ grows at late times,
since the $\chi$ mass increases as $\phi$ rolls away from the ESP.
However, if the number density of $\chi$ particles decreases due
to annihilation or decay into lighter modes, this mass
amplification effect is lost. It is therefore important to
determine the rate of decay and annihilation of the $\chi$
particles.

In Appendix B we address these issues and demonstrate
that the trapping effect is robust for certain parameter ranges,
provided that the light states are relatively stable.  This stability can easily
be arranged in supersymmetric models, and in fact
occurs automatically in certain D-brane systems.

Rescattering effects, in contrast, may actually strengthen the
trapping effect. Once $\chi$ particles have been created, they
will scatter off of the homogeneous $\phi$ condensate, causing it
to gradually decay into inhomogeneous, decoherent $\phi$
excitations \refs{\kls,\KhlebnikovMC,\FelderHQ}. However, we will not
consider this potentially beneficial effect here.

\subsec{The Example of Moving D-branes}

Before proceeding, it may be illustrative to discuss these results
in terms of a simple, mechanical example -- a moving pair of
D-branes. The moduli space of a system of two D-branes is the
space of brane positions.  In terms of the brane worldvolume
fields the separation between the two branes can be regarded as a
Higgs field $\phi$. The off-diagonal components of the $U(2)$
gauge field are the $W$ bosons.  At the ESP of this system,
$\phi=0$, the $W$ bosons are massless. Away from $\phi=0$ the $W$
bosons acquire a mass by the Higgs mechanism, breaking the
symmetry group from $U(2)$ down to $U(1)\times U(1)$. If we
identify $\chi$ with the $W$ field\foot{For simplicity we ignore
the superpartner of the $\chi$ boson.} and $g^2 \sim g_{YM}^2 \sim
g_s$ with the string coupling, then we find that the brane
worldvolume theory contains a term like \lis.  We therefore expect
this system to exhibit moduli trapping.

\global\advance\figno by 1

\ifig\branes{This figure illustrates the creation of open strings
as two D-branes pass near each other. The left corner
 shows the target space picture of the creation of the open strings.}
{\epsfxsize4in\epsfbox{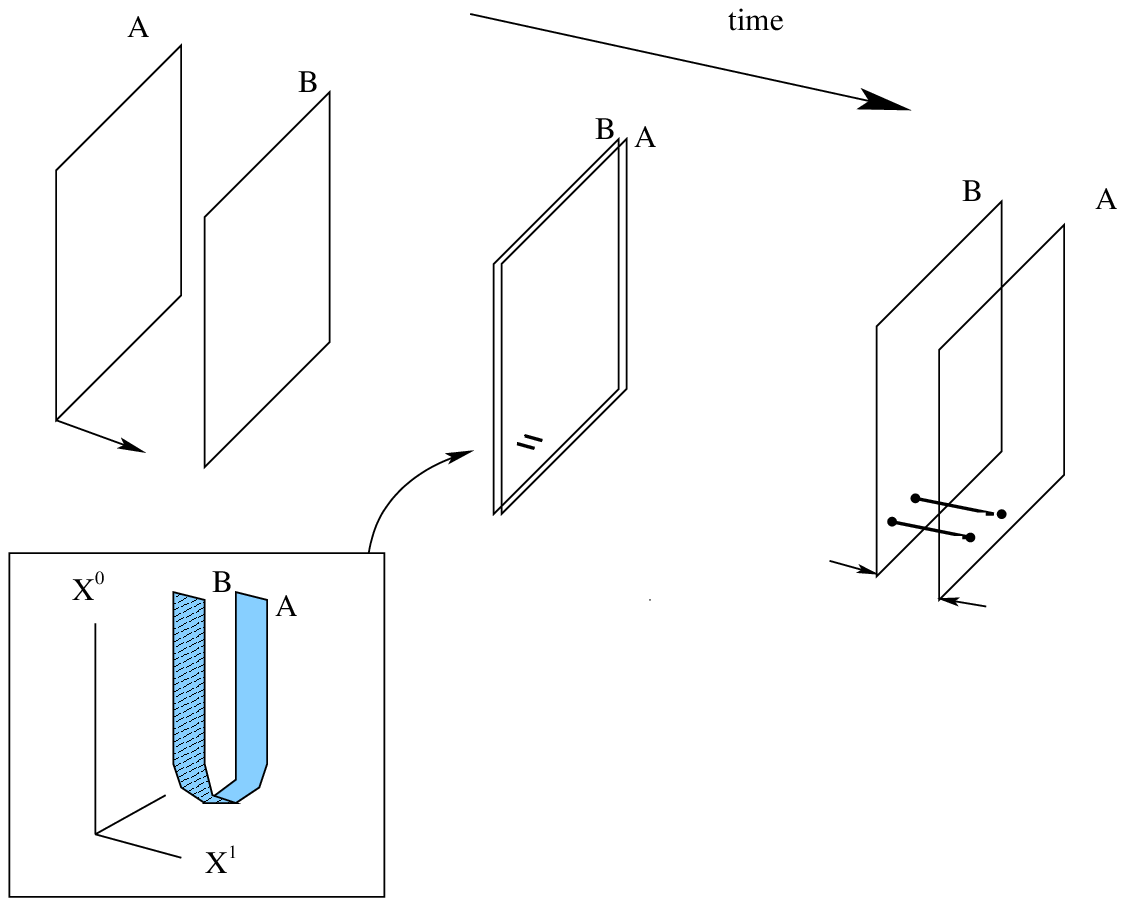}}

The trapping effect is a quantum correction to the motion of
D-branes. As the D-branes approach each other, the open strings
stretched between them become excited. When the D-branes pass by
each other and begin moving apart the stretched open strings
become massive and pull the D-branes back together.  We depict
this in \branes.

This effect can be a significant correction to the dynamics of any
system with a number of mobile, mutually BPS D-branes.  Consider,
for example, $N$ D3-branes which fill spacetime and are transverse
to a compact six-manifold $M$.  Let us take these branes to begin
with small, random, classical velocities in $M$. The classical
dynamics of this system is similar to that of a nonrelativistic,
noninteracting, classical gas. When we include quantum production
of light strings, the branes begin to trap each other, pairwise or
in small groups, then gradually agglomerate until only a few
massive clumps of many branes remain.

 One interesting consequence is that such a system will tend to exhibit
enhanced gauge symmetry, with gauge group $U(N)$ if the final
state consists of a single clump.  (Hubble friction may bring the
branes to rest before the aggregation is complete, in which case
the gauge group will be a product of smaller factors; we will
address related issues in \S4.1.) Another important effect of
massive clumps is their gravitational backreaction: a large
cluster of D-branes will produce a warped throat region in $M$,
which may be of phenomenological interest \rs.

 There are additional corrections to the classical moduli space
approximation of the D-brane motion which come from
velocity-dependent forces. These correspond in the D-brane
worldvolume field theory to higher-derivative corrections
generated by virtual effects.  When this field theory is at weak
't Hooft coupling, open string production is the dominant effect
as one approaches an ESP.  However, sufficiently large clusters of
branes will be described by gauge theories at strong 't Hooft
coupling, where the dynamics of additional probe branes is
governed instead by the analysis of \SilversteinHF.\foot{A further
correction to our dynamics could arise if, as we will discuss in
\S5, the branes keep moving until the system is beyond the range
of effective field theory.}

A similar interaction was studied in the context of the scattering
of D0-branes in \DouglasYP.  There is a crucial difference between
that system and the case of interest here, in which the branes are
extended along $3+1$ dimensions.  In the D0-brane problem, there
is a nontrivial probability for the D0-branes to pass by each
other without getting trapped: because the D0-brane is pointlike,
there is some probability for no open strings between them to be
created or for those created to annihilate rapidly.  This is the
leading contribution to the S-matrix.  In our case, there is
always a nonzero number {\it density} of particles created. As we
argue in Appendix B, for certain ranges of parameters these
particles do not annihilate rapidly enough to prevent trapping.

\newsec{Moduli Trapping: Detailed Analysis}

In the previous section we gave an intuitive explanation of the
trapping effect, which we will now describe in more detail. In
\S3.1 we will present the equations of motion which govern the
trajectory of the modulus $\phi$, including the backreaction due
to production of light particles.  These equations are difficult
to solve exactly, so in \S3.2 we will integrate the system
numerically. In \S3.3 we focus on the special case $\mu=0$, where
the modulus rolls directly through the ESP. In this case analytic
techniques are available, and as we will see the trapping effect
is considerably stronger than in the  $\mu\ne 0$ case. Even for
$\mu \ne 0$ some aspects can be derived analytically in a
perturbative expansion. This last case is somewhat technical, so
we refer the reader to Appendix C for details.

\subsec{Formal Description of Particle Production Near an ESP}

The full equations of motion are found by coupling the classical
motion of $\phi$ to the time-dependent $\chi$ quantum field theory
defined by \lis.\foot{We remain in flat space quantum field
theory, reserving gravitational effects for \S4.}

In general, the presence of an ESP will alter the moduli dynamics
in two ways. First, any $\chi$ excitations produced by the
mechanism described above will backreact on the classical
evolution of $\vphi$.  In particular, as we saw in
\endens, a non-zero expectation value $\la \chi^2 \ra \ne 0$
arising from particle production effectively acts like a
linear potential for $\vphi$ and drives the moduli towards the
origin. This is the effect we wish to describe. Second, virtual
$\chi$ particles generate quadratic and higher-derivative
contributions to the effective action as well as an effective
potential for a spacetime-homogeneous $\vphi$.

As we discussed in \S2.1, we can neglect the kinetic corrections
in our weakly-coupled situation. The interaction in \lis\ also
induces important radiative corrections to the effective
potential. Specifically, it leads to a Coleman-Weinberg effective
potential and three UV-divergent terms:
\eqn\veff{ V_{eff}(\vphi) = \Lambda_{eff} + g^2 m^2_{eff}  \vphi^2
+ g^4 \lambda_{eff}\vphi^4 .}
These UV divergences could be subtracted by hand using appropriate
counterterms. In a supersymmetric system these divergences are
absent.

In order to isolate the effects of particle production at the
order  we are working, we will subtract by hand the entire
Coleman-Weinberg effective potential for $\phi$ that is generated
by one loop of $\chi$ particles. This mimics the effect of
including extended supersymmetry, which is a toy case of interest
in string theory and supergravity. For the more realistic ${\cal
N}=1$ supersymmetry in four dimensions, radiative corrections do
generically generate a nontrivial potential energy.  Nevertheless,
particle production effects can still dominate the virtual
corrections to the potential after spontaneous supersymmetry
breaking.  The reason is that bosons and fermions contribute with
opposite signs in loops, but on-shell bosons and fermions, such as
those produced by the changing mass of $\chi$, contribute
with the same sign to backreaction on $\phi$.

To describe the production of $\chi$ particles, we first expand
the quantum field $\chi$ in terms of Fock space operators as
\eqn\fock{ \chi = \sum_k a_k \chi_k + a^\dagger_k \chi^*_k } where
the $\chi_k$ are a complete set of positive-frequency solutions to
the Klein-Gordon equation with mass \eqn\asd{m_\chi^2(t) = g^2
|\vphi(t)|^2.} Expanding in plane waves
\eqn\modes{ \chi_k = u_k(t) e^{ i k \cdot x}}
the equation of motion is
%
\eqn\ueq{ \Bigl(\p_t^2 + k^2 + g^2|\phi(t)|^2\Bigr)u_k = 0 .}
The modes \modes\ are normalized with respect to the Klein-Gordon
inner product, which fixes
\eqn\norm{
u^*_k {\dot u}_k - {\dot u}^*_k u_k = -i .}
The wave equation \ueq\ has two linearly-independent solutions for
each $k$, so in general there will be many inequivalent choices of
positive-frequency modes $\chi_k$. Each such choice of mode
decomposition defines a set of Fock space operators via \fock,
which in turn define a vacuum state of the theory. The wave
equation depends explicitly on time, so there is no canonical
choice of Poincar\'{e} invariant vacuum. Instead, there is a large
family of inequivalent vacua for $\chi$.

We can choose a set of positive frequency modes $u^{in}_k$ that
take a particularly simple form in the far past,
\eqn\inwaveftn{ u_k^{in}\to {1\over \sqrt{2\sqrt{k^2+g^2|\phi|^2
}}}e^{-i\int^t \sqrt{k^2+g^2|\phi(t^\prime)|^2 }dt^\prime} ~~~~~~
{\rm as }~~ t\to -\infty.}
%
This choice of mode
decomposition  defines a vacuum state $|in \rangle$.  In the far
past the phases of the solutions \inwaveftn\ are monotone
decreasing with $t$, indicating that the state $|in\rangle$ has no
particles in the far past.
This state, known as the adiabatic vacuum, evolves into a highly excited
state as the modulus $\vphi$ rolls past the ESP.

We can now write down the classical equation of motion for $\vphi$
including the effects of $\chi$ production. Including a
subtraction $\delta_M$, to be determined shortly, it is

\eqn\eomphi{ \Bigl(\p^2 + g^2 (\la \chi^2 \ra-\delta_M)\Bigr)\vphi
=0.}
The expectation value $\la \chi^2 \ra$ depends on time and is
calculated in the adiabatic vacuum $|in\ra$.  At time $t$
\eqn\wis{\eqalign{ \la in| \chi^2 (t) |in\ra &= \int {d^3{k} \over
(2\pi)^3}|u^{in}_k (t)|^2}.} where the $u^{in}_k$ are determined
by the boundary condition \inwaveftn\ in the far past.

In order to subtract the Coleman-Weinberg potential, we must
remove the contribution to $\la \chi^2 \ra$ coming from one loop
of $\chi$ particles, replacing the $\chi$ mass-squared with
$g^2|\phi(t)|^2$. That is, the subtraction $\delta_M$ can be
written as
\eqn\mcounter{\delta_M\equiv \int
{d^3{k}\over(2\pi)^3}{1\over{2\sqrt{k^2+g^2|\phi|^2}}} .}
With this form it is straightforward to see that when the impact
parameter is very large, $(\la \chi^2 \ra - \delta_M)$ is
negligible and $\vphi$ follows its original trajectory \phiis.

To summarize, the effects of quantum production of $\chi$
particles on the classical motion of the modulus $\vphi$ are
governed by: \eqn\summary{\eqalign{\Bigl(\p^2 + g^2 (\la \chi^2
\ra-\delta_M)\Bigr)\vphi &=0 \cr \Bigl(\p_t^2 + k^2 +
g^2|\phi(t)|^2\Bigr)u^{in}_{k} &= 0  \cr \la \chi^2(t)\ra =
\int{{{d^3{k}\over{(2\pi)^3}}|u^{in}_{k}(t)|^2 }}&.}}

The above equations of motion can be reformulated in terms of the
energy transferred between the two systems. In particular, it is
straightforward to show that the coupled equations \summary\ are
equivalent to the statement \eqn\ham{ {d\over dt} H_\vphi = -
{d\over dt}\la in| H_{\chi} |in\ra .} The left-hand side of \ham\
involves the classical energy of the rolling $\vphi(t)$ fields,
whereas the right hand side is an expectation value of the
time-dependent $\chi$ Hamiltonian calculated in quantum field
theory. This is the more precise form of energy conservation which
applies to our rough estimate in \S2.2.

Furthermore, the angular momentum on moduli space is conserved,
since the action \lis\ is invariant under phase rotations $\phi
\to \phi e^{i\theta}$. In the present case \lis, the $\chi$
particles do not carry angular momentum, so the orbit of $\phi$
around the ESP will have fixed angular momentum.  The result is an
angular momentum barrier which keeps the modulus at a finite
distance from the ESP.

More complicated scenarios allow for the exchange of angular
momentum between $\phi$ and $\chi$.  This includes the case of
colliding D-branes, where the strings stretching between the two
D-branes can carry angular momentum.  Moreover, as we will see in
\S4, the situation changes once gravitational effects are
included, as angular momentum is redshifted away by cosmological
expansion. This leads to scenarios where the moduli are trapped
exactly at the ESP, rather than orbiting around it at some finite
distance.

\subsec{Moduli Trapping: Numerical Results}

The coupled set of integral and differential equations \summary\
governing the trapping trajectory is hard to solve in general.
Some analytic results can be obtained through an expansion in the
non-adiabaticity parameter $\dot \omega/\omega^2$, combined with a
systematic iteration procedure. Specifically, to the extent that
this non-adiabaticity parameter is small enough that the
correction to the motion of the moduli field only shows up at
exponentially late times, we can calculate the bending angle and
the energy loss during its first pass. This is given in Appendix
C.

As time goes on, the mass amplification of the $\chi$ particles
makes higher order terms as well as non-perturbative terms in the
adiabatic expansion crucial for the motion of the moduli. This
makes it very hard to proceed analytically to obtain the detailed
evolution of the system.

We have numerically integrated the coupled equations \summary\ in
Mathematica, using a discrete sum to approximate the momentum
integral $k$, and implementing the subtraction of the
Coleman-Weinberg potential described above.

\ifig\trap{This figure shows the evolution, in the complex  $\phi$
plane, of a system with parameters $g^2=20,\mu=0.3,v=1$. The field
rolls in from the right and gets trapped into the precessing
orbit exhibited in the plot.  The orbit is initially
an elongated ellipse, but gradually becomes more circular. In an expanding
universe, the field would lose its angular momentum, so that the radius of
the circle would eventually shrink to zero.}
{\epsfxsize3.5in\epsfbox{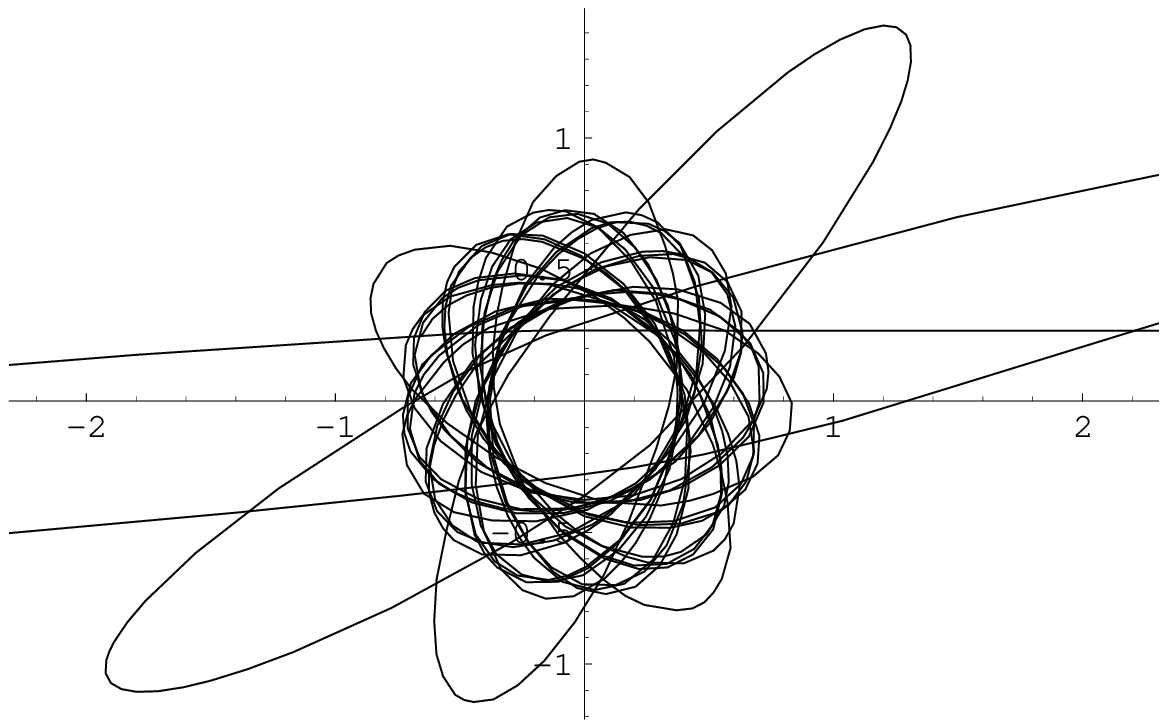}}

In \trap\ we plot a trajectory for the case $\mu >0$, where
$\phi$ becomes trapped in a spiral orbit around the ESP. The
radius of the orbit varies with the parameters, but the
qualitative features shown are typical.

\ifig\realtrap{This shows one-dimensional trapping, in which
$\phi$ passes directly through the ESP $\phi =0$.  The vertical
axis is the real part of $\phi$, and the  horizontal axis is time.
The amplitude of the oscillations decreases exponentially as a
result of parametric resonance, as we explain in \S3.3.}
{\epsfxsize3.2in\epsfbox{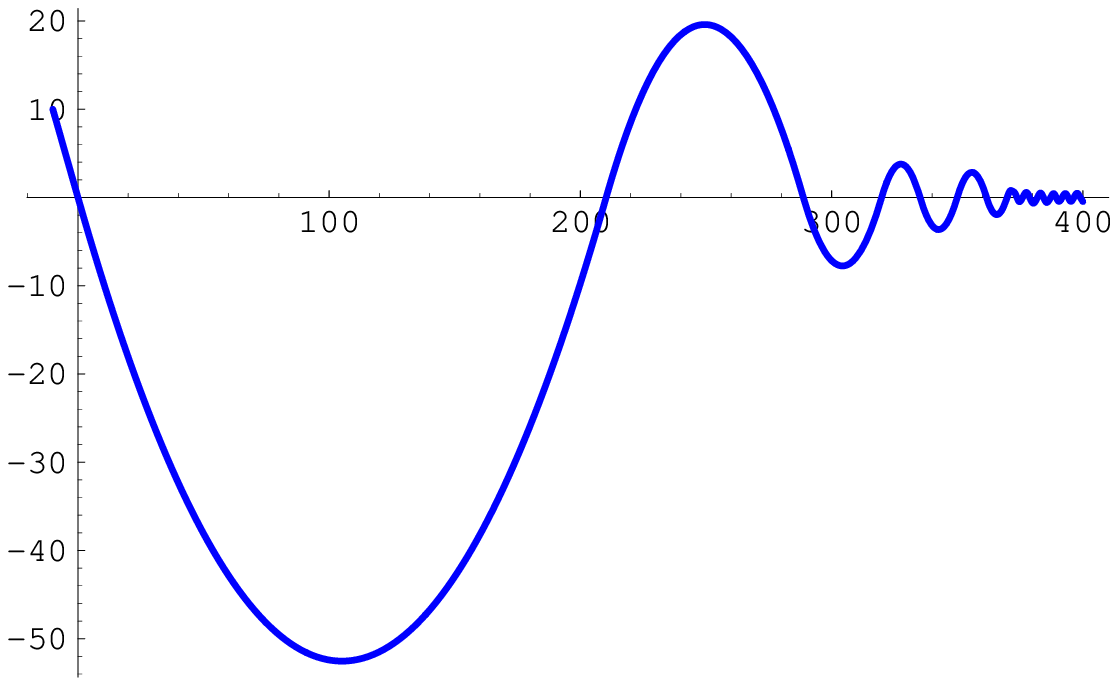}}

In \realtrap\ we plot the trajectory of a modulus which is aimed
to pass directly through an ESP, with vanishing impact parameter.
In this case the motion becomes effectively one-dimensional, and
the field moves directly through the ESP $\phi =0$. The trapping
effect in this case is especially strong, and can be understood
analytically to come from resonant production of $\chi$ particles,
as we will now explain.

\subsec{The Special Case of One-Dimensional Motion}

In this section we will concentrate on the interesting and
important  special case of one-dimensional motion, i.e. vanishing
impact parameter $\mu$. Perhaps surprisingly, this is a good
approximation to the general case.  Indeed, the results of \S2
demonstrate that trapping becomes exponentially suppressed when
the impact parameter $\mu$ (the imaginary part of the moduli
field) becomes greater than $\sqrt{v\over \pi g}$. On the other
hand, for $\mu \ll \sqrt{v\over \pi g}$ the motion of the field
$\phi$ stops at $\phi_* \sim {{4 \pi^3}v^{1/2}\over{g^{5/2 }}}$.
The ratio of $\phi_*$ to $\mu$ in the regime where trapping is
efficient (i.e. for $\mu < \sqrt{v\over \pi g}$) is therefore
\eqn\ellipticity{{\phi_*\over\mu} > {4\pi^{7/2}\over g^2} \ .}
Thus, in the case of efficient trapping and weak coupling, the
ellipticity of the moduli orbit is very high, so that the motion
is effectively one-dimensional.

In the case $\mu=0$ the number density of $\chi$ particles created when
the field $\phi$ passes the ESP is
\eqn\numdens{n_{{}_{\chi}}={{(gv)^{3 \over 2}}\over{(2\pi)^3}} \ .
}
At $|\phi| \gg \sqrt{v\over g}$, when the $\chi$ particles are
nonrelativistic, the mass of each particle is equal to $g |\phi|$,
and their energy density is given by \instant
\eqn\endensII{\rho_{{}_{\chi}}(\phi)= g n_{{}_{\chi}} |\phi| =
{{g^{5 \over 2}v^{3/2}}\over{(2\pi)^3}}~|\phi|}
We have written $|\phi|$ because this energy does not depend
on the sign of the field $\phi$. This will be very important for
us in what follows.

One should note that, strictly speaking, the $\chi$ particles have
some kinetic energy even at $\phi =0$, but for $g \ll 1$ this
energy is much smaller than the kinetic energy of $\phi$
\instant: \eqn\rhokin{ \rho_{{}_{\chi}}(\phi = 0) \sim {g^2 \over
4\pi^{7/2}} ~{v^2\over 2} = {g^2\over 4\pi^{7/2}}~ \rho^{\rm
kin}_{\phi} . } This means that the energy of $\phi$
decreases only slightly when it passes through the ESP $\phi = 0$.
Although the initial energy in $\chi$ particles is small, this
energy increases with $|\phi|$, $\rho_{{}_{\chi}}\sim g
n_{{}_{\chi}} |\phi|$, and creates an effective potential for
$\phi$. The equation of motion for $\phi$ in this potential is
\kls: \eqn\uu{ \ddot \phi + g {n_{{}_{\chi}}} {\phi\over|\phi|} =
0\ . } The last term means that $\phi$ is attracted to the ESP
$\phi =0$ with a constant force proportional to $n_{{}_{\chi}}$.

At some location $\phi_{1}^*$ the $\chi$ energy density
$\rho_{{}_{\chi}}$ equals the initial kinetic energy density
${1\over 2}\dot\phi^2\equiv {1\over 2} v^2$; at this point $\phi$
stops and then falls back toward $\phi=0$.

On this second pass by the origin, the energy density of the
$\chi$ particles again becomes much smaller than the kinetic
energy of $\phi$.  Energy conservation implies that $\phi$ will
pass the point $\phi = 0$ at almost exactly the initial velocity
$v$.  Since the conditions are almost the same as on the first
pass, new $\chi$ particles will be created, i.e. $n_{{}_{\chi}}$
will increase. The field $\phi$ will continue moving for a while,
stop at some point $\phi^*_2$, and then fall back once more to the
ESP, creating more particles. Because each new collection of
particles is created in the presence of previous generations of
particles, the process occurs in the regime of parametric
resonance, as in the theory of preheating.

A detailed theory of this process was considered in \kls; see in
particular Eqs. (59),(60). By translating the problem into a
one-dimensional quantum mechanics system (as in Appendix A) with a
particle scattering repeatedly across an inverted harmonic
potential, \kls\ calculated the multiplicative increase of the
Bogoliubov coefficients during each pass in terms of the
reflection and transition amplitudes. In application to our
problem, the equations describing the occupation numbers of $\chi$
particles with momentum $k$ produced when the field passes through
the ESP $j+1$ times look as follows:
\eqn\index{ n^{j+1}_k
=n^{j}_k \exp( 2\pi \mu_k^{j}),}
where
\eqn\muj{\mu_k^{j}={1 \over 2\pi} \ln \left( 1 +2 e^{-\pi \xi^2}
 - 2\sin \theta^{j}~ e^{-{{\pi \over2}
 \xi^2}}~\sqrt{1+ e^{-\pi \xi^2}} \right).} %
Here $\xi^2 = {k^2\over{g v}}$ and $\theta^{j}$ is a relative
phase variable which takes values from $0 $ to $2\pi$. In a cyclic
particle creation process in which the parameters of the system
change considerably during each oscillation (which is our case, as
will become clear shortly), the phases $\theta^{j}$ change almost
randomly. As a result, the coefficient $\mu_j$ for small $k$ takes
different values, from $0.28$ to $-0.28$, but for 3/4 of all
values of the angle $\theta^j$ the coefficient $\mu_j$ is
positive.  The average value of $\mu_j$ is approximately equal to
$0.15$. This means that, on average, the number density of $\chi$
particles grows by approximately a factor of two or three each
time that $\phi$ passes through the ESP $\phi = 0$.

But this means that with each pass, the coefficient
$n_{{}_{\chi}}$ in \endensII\ grows by a factor of two or three.
It follows that the effective potential becomes two to three times
more steep with each pass. Correspondingly, the maximal deviation
$|\phi^*_i|$ from the point $\phi=0$ exponentially decreases with
each new oscillation. Since the velocity of the field at the point
$\phi=0$ remains almost unchanged until $\phi$ loses its energy to
the created particles, the duration of each oscillation decreases
exponentially as well. Therefore the whole process takes a time
${\cal{O}}(10)\phi_1^*/v$, after which the backreaction of the
created particles becomes important, and the field falls to the
ESP.

This process is very similar to the last stages
of preheating, as studied in \kls. The main difference is that in the
simplest models of preheating the field oscillates near the
minimum of its classical potential. In our case the effective
potential is initially absent, but a potential is generated due to the
created
particles. This is exactly what happens at the late stages of
preheating, when the effective potential (with an account taken of
the produced particles) becomes dominated by the rapidly-growing
term proportional to $|\phi|$; see the discussion in Section VIII
B of \kls.

We would like to emphasize that until the very last stages of the
process, the backreaction of the created particles can be studied
by the simple methods described above. At this stage the total
number of created particles is still very small, but their number
grows exponentially with each new oscillation. This leads to an
exponentially rapid increase of the steepness of the potential
energy of the field $\phi$
\endensII\ and, correspondingly, to an exponentially rapid
decrease of the amplitude of its oscillations. This extremely fast
trapping of $\phi$ happens despite the fact that at this first
stage of oscillations the total energy of $\phi$, including its
potential energy, remains almost constant.

Once the amplitude of oscillations becomes smaller than the width
of the nonadiabaticity region, $|\phi(t)| \lesssim \Delta \phi
\sim \sqrt{v/g}$, one can no longer assume that the number of
particles will continue to grow via a rapidly-developing
parametric resonance. The amplitude of the oscillations is given
by ${v^2\over 2g|\phi| n_\chi}$, so the amplitude becomes
${\cal{O}}(\sqrt{v/g})$ when the total number of the produced
particles grows to
\eqn\numberw{ n_{\chi} \sim v^{3/2} g^{-1/2} \ .}
Note that the typical energy of each  $\chi $ particle at
$|\phi(t)| \sim \sqrt{v/g}$ is of the same order as its kinetic
energy ${\cal{O}}(\sqrt{gv})$. One can easily see that the total energy
density of particles $\chi$ at that stage is roughly $ \sqrt{gv}
n_\chi \sim {\cal{O}}(v^2)$, i.e. it is comparable to the initial kinetic
energy of $\phi$.

Thus, our estimates indicate that the regime of the broad
parametric resonance ends when a substantial part of the initial
kinetic energy of $\phi$ is converted to the energy of
the $\chi$ particles, and the amplitude of the oscillating field
$\phi$ becomes comparable to the width of the nonadiabaticity
region,
\eqn\finalampl{|\phi| \sim \Delta\phi = \sqrt{v/g} \ .}

We will use these estimates in our discussion of the cosmological
consequences of moduli trapping. In order to obtain a more
complete and reliable description of the last stages of this
process one should use lattice simulations, taking into account
the rescattering of created particles \refs{\KhlebnikovMC,\FelderHQ}. An
investigation of a similar situation in the theory of preheating
has shown that rescattering makes the process of particle
production more efficient. This speeds up the last stages of
particle production and leads to a rapid decay of the
field $\phi$ \FelderHR, which in our case corresponds to a rapid
descent of $\phi$ toward the enhanced symmetry point.

\newsec{Trapped Moduli in an Expanding Universe}

\subsec{Rapid Trapping}

In this section we will study the conditions under which the
trapping mechanism in quantum field theory survives the effects of
coupling to gravity in an expanding universe.

First, we should point out one very beneficial effect of
cosmological expansion. The field-theoretic mechanism presented
above often leads to moduli being trapped in large-amplitude
fluctuations \matchen\ around an ESP when $\mu\ne 0$. On
timescales where the expansion is noticeable, Hubble friction will
naturally extract the energy from this motion, drawing the modulus
inward and leading the modulus to come to rest at the ESP.

Let us now ask whether the expansion of the universe can impede
moduli trapping. Consider a system of moduli coupled to gravity,
with the fields arranged to roll near an ESP.  For simplicity we
will consider FRW solutions with flat spatial slices,
\eqn\frw{ds^2=dt^2-a(t)^2d\vec x^2.}
The Friedman equation determining $a(t)$ is
\eqn\friedeq{3H^2={1\over M_p^2}\rho}
where $H=\dot a/a$ and $\rho$ is the energy density of the moduli.

The trapping effect will be robust against cosmological expansion
if the timescale governing trapping is short compared to $H^{-1}$,
i.e. if $H \ll v/\phi_* $, where $\phi_{*}$ is given by \matchen.
Assuming that the potential energy of the moduli is non-negative,
this implies that

\eqn\nearcond{ \phi_{*} \ll \sqrt{6} M_{p} \to
{{{4 \pi^3}\over{6^{1/2}g^{5/2}}}{{v^{1/2}\over{M_p}}}e^{\pi g
\mu^2/v}} \ll 1 .}
This condition suffices to ensure that trapping is very rapid.

If this condition is satisfied, trapping occurs in much less
than  a Hubble time, in which case the analysis of \S2 and \S3
remains valid. We will show in \S4.3 that even when \nearcond\ is
not satisfied, trapping does still occur, although with somewhat
different dynamics.

\subsec{Scanning Range in an Expanding Universe}

An important effect of the gravitational coupling is that during
the expansion of the universe, the energy density in produced
$\chi$ particles dilutes like $1/a^3$ if they are non-relativistic
and like $1/a^4$ if they are relativistic.  The energy in coherent
motion of $\phi$, however, has the equation of state $p = \rho$
and therefore dilutes much faster, as $1/a^6$.

This effect reduces the range of motion for the moduli even before
they encounter any ESPs. Hubble friction slows the progress of any
rolling scalar field, and if the distance between ESPs is
sufficiently large then a typical rolling modulus will come to
rest without ever passing near an ESP.  In order to apply our
results to the vacuum selection problem, we will need to know how
large a range of $\phi$ we can scan over in the presence of Hubble
friction. This can be obtained as follows \FelderJK.

If we are in an FRW phase, \eqn\frw{ a(t)=a_{0} t^{\beta}} then
the equation of motion for $\phi$ (ignoring any potential terms)
\eqn\eom{\ddot{\phi}+3 H \dot\phi = 0} has solutions of the form
\eqn\sol{\dot{\phi}(t) = v \left({t_0\over{t}}\right)^{3\beta}.}
We can integrate this to determine how far the field rolls before
stopping.

Let us first consider the case $\beta = 1/3$, which corresponds to
the equation of state $p = \rho$.  This includes the case where
the coherent, classical kinetic energy of $\phi$ drives the
expansion.  The value of $\phi$, \eqn\distanceone{ \phi(t) = v
t_{0}~{\rm{log}}\left({t\over{t_0}}\right)} diverges at large $t$.
Thus $\phi$ can travel an arbitrarily large distance in moduli
space.

In the more general case $\beta > 1/3$ the field will travel a distance

\eqn\distanceis{\phi(t)-\phi(t_0) = {v\over{H(t_{0})}}
{\beta\over{3\beta-1}}} before stopping.

In order to be in a phase with $\beta > 1/3$, the kinetic energy
of  $\phi$ must not be totally dominant; that is, we must have
${1\over{2}}\dot\phi^2<\rho$, where $\rho \equiv 3M_p^2 H^2$ is
the total energy density appearing on the right hand side of the
Friedman equation. Plugging this into \distanceis\  we obtain the
constraint
\eqn\scanbound{\phi(t)-\phi(t_0)<\sqrt{6}M_p{\beta\over{3\beta-1}}.}

Let us consider a specific example.  Suppose that we start at
$t_0$ with kinetic energy domination: $K_0/\rho_0 = 1 -\epsilon$,
$\epsilon \ll 1$, in some region of the universe that can be
modelled as an expanding FRW cosmology. The kinetic energy drops
like $K \sim \rho_0(a_0/a)^6 \sim \rho_0 (t_0 / t)^2$, while the
other components of the energy dilute like
\eqn\otherdilute{\rho(t) = \epsilon \rho_0 (t_0/t)^{1 + w},}  with
$w<1$. The universe will stop being kinetic-energy dominated at
the time $t_c = t_0 {\epsilon}^{- 1/{(1-w)} }$, at which point,
according to \distanceone, the modulus has travelled a distance
\eqn\distone{\phi(t_c) - \phi(t_0) = - {1 \over {1 - w}}v t_0 \log
\epsilon .} After this the field keeps moving and covers an
additional range \eqn\disttwo{\phi(t_*)- \phi(t_c) = \sqrt{3} M_p
{2 \over {3(1 - w )}}.}

To get a feel for the numbers, consider the case where $ vt_0 \sim
M_{p}$, $\epsilon \sim 10^{-2}$, and $w = 0$. Then $\phi$ will
travel a total distance $\phi(t_*) - \phi(t_0)  \sim 6 M_p$ in
field space, which is not particularly far. However, as we will
discuss in \S6, certain moduli spaces of interest have a rich
structure on sub-Planckian scales, so in these cases there is a
good chance that the modulus will encounter an ESP and get trapped
before Hubble friction brings the system to rest.

There is another natural possibility if we assume low-energy
${\cal N}=1$ supersymmetry.  If the moduli acquire their
potentials from supersymmetry breaking then there is a large ratio
between the Planck scale and the scale of these potentials,
leading to significant scanning ranges. Specifically, consider a
contribution to the energy density coming from a potential energy
$V$ at the supersymmetry-breaking scale. If the initial kinetic
energy of the moduli is Planckian and the supersymmetry-breaking
scale is TeV then there will be a prolonged phase in which kinetic
energy dominates, since $\epsilon=V/M_p^4 \sim 10^{-64}$. This
allows $\phi$ to scan a significantly super-Planckian range in
field space.

\subsec{Trapping in an Expanding Universe}

We are now in a position to combine all the relevant effects and
consider trapping during expansion of the universe. For
simplicity, we will concentrate on the case of effectively
one-dimensional motion, $\mu \ll \sqrt{v/g}$. Suppose that, taking
into account Hubble friction, the modulus field passes in the
vicinity of the ESP at some moment $t_0$, so that $\chi$ particles
are produced, with $n_\chi(t_0) = {(gv)^{3/2}\over (2\pi)^3}$. We
will now determine the remaining evolution including both our
trapping force and Hubble friction. After the particles have
been produced, the field $\phi$ becomes attracted toward
$\phi =0$ by a force $gn_\chi$, so taking into account the
dilution of the produced particles, for $\phi
> 0$ the equation of motion is
\eqn\eomold{ \ddot\phi+3H\dot\phi=-g n_\chi(t_0)\left({a(t_0)\over
a(t)}\right)^3 }
For the general power law case, $a(t)\propto t^\beta$, this
becomes
\eqn\eomnew{ \ddot\phi+3{\beta\over t}\dot\phi=-g
n_\chi(t_0)(t_0/t)^{3\beta} \ .}
The general solution of this equation  is
\eqn\solnnew{\phi(t)=\phi(t_0)+c(t_0^{-3\beta+1}
-t^{-3\beta+1})+{g n_\chi(t_0)t_0^2\over(2-3\beta)}-{g
n_\chi(t_0)\over(2-3\beta)}(t_0/t)^{3\beta}t^2}
where $c$ is some constant. In the important case $\beta = 2/3$,
which corresponds to a universe dominated by pressureless cold
matter, the general solution is
\eqn\zeropressure{ \phi = \phi(0) + c(t_0^{-1} -t^{-1})-g n_\chi
t_0^2 \log{t\over t_0} \ .}
According to these solutions, in a universe dominated by matter
with non-negative pressure (i.e. $\beta \le 2/3$) the field
$\phi$ moves to $-\infty$ as $t\to\infty$.

Of course, as soon as the field reaches the point $\phi=0$, this
solution is no longer applicable, since the attractive force
changes its sign (the potential is proportional to $|\phi|$). The
result above simply means that the attractive force is always
strong enough to bring the field back to the point $\phi = 0$
within finite time. Then the field moves further, with ever
decreasing speed, turns back again, and returns to $\phi = 0$ once
again. The amplitude of each oscillation rapidly decreases due to
the combined effect of the Hubble friction and of the (weak)
parametric resonance. This means that once $\phi$ passes near the
ESP, its fate is sealed: eventually it will be trapped there.

\subsec{Efficiency of Trapping}

It is useful to determine what fraction of all initial conditions
for the moving moduli lead to trapping. There are several
constraints to be satisfied. First of all, if the impact parameter
$\mu$ is much larger than $\sqrt{v/g}$, the number of produced
particles will be exponentially small, and the efficiency of
trapping will be exponentially suppressed. Of course, eventually
$\phi$ will fall to the enhanced symmetry point, but if this
process takes an exponentially large time, the trapping effect
will be of no practical significance. Thus one can roughly
estimate the range of interesting impact parameters to be
${\cal{O}}(\sqrt{v/g})$.

Another constraint is related to the fact that even if initially
the energy density of the universe was dominated by the moving
moduli, as discussed in \S4.2, these fields can only move the
distance given by \distone,\disttwo. This distance depends on the
initial ratio $1-\epsilon$ of kinetic energy to total energy, leading to a
scanning range $C M_p$ in field space, where the prefactor $C$ is
logarithmically related to $\epsilon$.

Thus, the field becomes trapped only if there is an enhanced
symmetry point inside a rectangle with sides of length $C M_p$
along the direction of motion and width ${\cal{O}}(\sqrt{v/g})$ in
the direction perpendicular to the motion.

Interestingly, the total area (phase space) of the moduli trap
\eqn\traparea{S_{\rm trap} \sim  C M_p \sqrt {v\over g}}
increases as the coupling decreases.  This implies that the efficiency of
trapping grows at weak coupling.  Although
this may seem paradoxical, it happens because the mass of the
$\chi$ particles is proportional to the coupling constant and
(fixing the other parameters) it is easier to produce lighter
particles. On the other hand, if $g$ becomes too small, the
trapping force $gn_\chi\sim g^{5/2}v^{3/2}$ becomes smaller than
the usual forces due to the effective potential, which we assumed
subdominant in our investigation.

So far we have studied the simplest model where only one scalar
field becomes massless at the enhanced symmetry point. Let us
suppose, however, that N fields become massless at the point $\phi
= 0$. If these fields interact with $\phi$ with the same coupling
constant $g$, then particles of each of these fields are produced,
and the trapping force becomes N times stronger. In other words,
the trapping force is proportional to the degree of symmetry at
the ESP.

\newsec{String Theory Effects}

It is interesting to ask if there is any controlled situation
where string-theoretic effects become important for moduli
trapping. Here we will simply list several circumstances in which
stringy and/or quantum gravity effects can come into play, as well
as some constraints on these effects, leaving a full analysis of
this subtle and interesting situation for the future.

\subsec{Large $\chi$ Mass}

One way stringy and quantum gravity effects could become important
in the colliding D-brane case is if the $\chi$ mass at the
turnaround point is greater than string scale, $g\phi_*>m_s$.
This can happen even if the velocity is so small that during the
non-adiabatic period near the origin only unexcited stretched
strings are created. Then, as in our above field theory analysis,
we have
\eqn\bigphistar{g\phi_* = {4\pi^3\over g^{3/2}}v^{1/2}e^{\pi
g\mu^2/v} .}
In this case, the full system includes modes, namely the created
$\chi$ strings, which are heavier than the string oscillator mode
excitations on the individual branes. This means that the system
as a whole cannot consistently be captured by pure effective field
theory. However, it may still happen that the created stretched
strings are relatively stable against annihilation or decay into
the lighter stringy modes.  Their annihilation cross section is
suppressed by their large mass, as discussed in Appendix
B.\foot{For stringy densities of stretched strings, there could be
additional corrections to the annihilation rate, but we will not
consider this possibility.} Furthermore, an individual stretched
string will not directly decay if it is the lightest particle
carrying a conserved charge.

This latter situation happens in the simplest version of a D-brane
collision. The created stretched string cannot decay into lighter
string or field theory modes because it is charged and they are
not.

\subsec{Large $v$ and the Hagedorn Density of States}

If we increase the field velocity $\dot\phi=v$, then we may obtain
a situation in which excited string states are produced as $\phi$
passes the ESP.  The number of string states produced in this
process is enhanced by the Hagedorn density of states, so the
Bogoliubov  coefficients have the structure
\eqn\hagedornbeta{|\beta_k|^2=\sum_n e^{{\sqrt{n}\over
2\pi\sqrt{2}}}e^{-\pi(k^2+nm_s^2+g^2\mu^2
)/(gv)}}
where in the D-brane context, $g=\sqrt{g_s}$ is the Yang-Mills
coupling on the D-branes. Because of the $e^{-\pi n m_s^2/(gv)}$
suppression in the second factor, this effect is only significant
if $gv \gg m_{s}^2$.

However, in the case of colliding D-branes, and any situation dual to
it, there is a fundamental bound on the field velocity from the
relativistic speed limit of the branes.  That is, for large velocity
one must include the full Dirac-Born-Infeld Lagrangian for
$\phi$, which takes the form
\eqn\DBI{S=-{1\over {(2\pi)^3 g_s \alpha'^2}}\int d^4x
\sqrt{1-g^2{\dot\phi^2\over m_s^4}}.}
This action governs the nontrivial dynamics of $\phi$ for
velocities approaching the string scale, and in particular, it
reflects the fact that the brane velocity $g\dot\phi\alpha^\prime$
must be less than the speed of light in the ambient space. Applied
to our situation, \DBI\ implies that the D-brane velocity cannot
be large enough for the Hagedorn enhancement \hagedornbeta\ to
substantially increase the trapping effect.

However, in the presence of a large velocity, the effective mass
of the stretched string also has important velocity-dependent
contributions \SilversteinHF.  This may enhance the
non-adiabaticity near the origin and thus enhance the particle
production effect.  If anything, we expect this to increase the
trapping effect; it would be interesting to study this case
further.  As we discuss at the end of Appendix A, by using
unitarity combined with the stringy calculations in \BachasKX, it
might be possible to determine the net result of all these effects
on the stringy trapping mechanism. It would be interesting if,
taking into account all relevant quantum effects, a large
contribution to string production occurred in some controlled
setting, as suggested in \refs{\StromingerPC,\MaloneyCK,\gubser}.
If such an effect occurred for motion near an ESP, it would
apparently enhance the trapping effect and possibly even indicate
that $\phi$ slows down enormously before passing through the
origin, as occurred in the case studied in \SilversteinHF.
However, it would be important to check for control of the quantum
corrections to such a system.

\subsec{Light Field-Theoretic Strings}

A further possibility is to formally reduce the tension of strings by
considering strings in warped throats, strings from branes
partially wrapped on shrinking cycles, and the like.  In these
situations, the strings are essentially field-theoretic, though
string theory techniques such as AdS/CFT and ``geometric
engineering" of field theories may provide technical help in
analyzing the situation.

\newsec{The Vacuum Selection Problem}

We can now apply the ideas of the previous five sections to the
cosmology of theories with moduli.

A natural application of the moduli trapping effect is to the
problem of vacuum selection. One mechanism of vacuum selection is
based on the dynamics of light scalars during inflation. Moduli
fields experience large quantum fluctuations during inflation and
can easily jump from one minimum (or valley) of their effective
potential to another. It was suggested long ago that such
processes may be responsible, e.g., for the choice of the vacuum
state in supersymmetric theories \SUSY\ and for the smallness of
the cosmological constant \LindeDQ. The probability of such
processes and the resulting field distribution depends on the
details of the inflationary scenario and the structure of the
effective potential \LindeXX.

The mechanism that we consider in this paper is, in a certain
sense, complementary to the inflationary mechanism discussed
above. During inflation the average velocities of the fields are
very small, but quantum fluctuations tend to take the light scalar
fields away from their equilibrium positions. On the other hand,
after inflation, the fields often find themselves not necessarily near
the minima of their potentials or in the valleys corresponding to
the flat directions, but on a hillside.  As they roll
down, they often acquire some speed along the valleys, see e.g.
\noinflation. At this stage (as well as in a possible pre-inflationary
epoch) the moduli trapping mechanism may operate.

This mechanism may reduce the question of how one vacuum
configuration is selected dynamically out of the entire moduli
space of vacua  to the question of how one ESP is selected out of
the set of all ESPs. This residual problem is much simpler because
ESPs generically comprise a tiny subset of the moduli space.

\subsec{Vacuum Selection in Quantum Field Theory}

In pure quantum field theory, discussed in \S2, we saw that if a
scalar field $\phi$ is initially aimed to pass near an ESP, then
$\phi$ gets drawn toward the ESP and is ultimately trapped there.
This appears to be a basic phenomenon in time-dependent quantum
field theory: moduli which begin in a coherent classical motion
typically become trapped at an ESP. This leads to a dynamical
preference for ESPs.

In many of the supersymmetric quantum field theories that have
been studied rigorously \sw, the moduli space contains singular
points at which light degrees of freedom emerge. We have seen that
moduli can become trapped near these points given suitable initial
conditions.

\subsec{Vacuum Selection in Supergravity and Superstring Cosmology}

Compactifications of M/string theory which have a description as a
low energy effective supersymmetric field theory can have a
natural separation of scales: the string or Planck scale can be
much larger than the energy scales in the effective field theory
potential.  Thus, the intrinsically stringy effects of \S5\ are
unimportant in this limit.  On the other hand, the effects of
coupling to gravity given in \S4 continue to provide a crucial
constraint, as we will now discuss.

First of all, as in the case of pure quantum field theory, there
exist very instructive toy models with extended supersymmetry, for
which there is no potential at all on the moduli space.  For these
examples, in situations where higher-derivative corrections to the
effective action are suppressed, a rolling scalar field has the
equation of state $p=\rho$. This corresponds to the $\beta=1/3$
case \distanceone\ of \S4, for which one can scan an arbitrarily
large distance in field space.  Therefore, in this case, the
trapping effect applies in a straightforward way to dynamically
select the ESPs for regimes in which \nearcond\ is also satisfied.

More generally, however, one may wish to implement cosmological
trapping in theories with some potential energy.  In this case the
requirement that the scanning range of $\phi$ (as constrained by
Hubble friction in \S4) should be large enough to cover multiple
vacua is an important constraint.  The absolute minimum
requirement is that the scanning range is sufficient for the
moduli to reach one ESP before stopping from Hubble friction; but
to address the vacuum selection problem one should ideally scan a
number of ESPs.

One context in which this can happen is in a phase in which the
kinetic energy of the rolling scalar fields dominates the energy
density of the universe so that the $\beta=1/3$ result
\distanceone\ applies. This may occur in a pre-inflationary phase
in some patches of spacetime, though it is subject to the
stringent limitation in duration given in \distone. Given such a
phase, the field will roll around until it gets trapped at an ESP.

During the ordinary radiation-dominated ($\beta=1/2$) and
matter-dominated ($\beta=2/3$) eras, the more stringent constraint
\scanbound\ applies.  As we indicated in \S4, this scanning range
is not large in Planck units, so we can usefully apply moduli
trapping to the problem of vacuum selection in these eras only if
the vacuum has appropriately rich structure on sub-Planckian
scales.  In other words, the average distance in moduli space
between ESPs should be sub-Planckian.

Gravitationally-coupled scalars $\phi$ generically have a
potential  energy $V(\phi/M_p)$ which has local minima separated
by Planck-scale distances.  In this cases, the limited scanning
range during the $\beta\ne 1/3$ cosmological eras prevents our
mechanism from addressing the vacuum selection problem. However,
it is generic for compactification moduli to have special ESPs
where the gravitationally-coupled system is enhanced to a system
with light field theory degrees of freedom. Given a rich enough
effective field theory in this ESP region, there will generically
be interesting vacuum structure on sub-Planckian distances.  In
this sort of region moduli trapping will pick out the ESP vacua of
the system.

\subsec{Properties of the Resulting Vacua }

Let us now consider the qualitative features of the vacua selected
by moduli trapping, assuming that the constraint imposed by Hubble
friction has been evaded in one of the ways described above.

First of all, it is important to recognize that what we have called ESPs may
well be subspaces of various dimensions, not points.  For example,
in toroidal compactification of the heterotic string, there is one
enhanced symmetry locus for each circle in the torus -- new states
appear when the circle is at the self-dual radius. Each of these
loci is codimension one in the moduli space, but of course their
intersections, where multiple radii are self-dual, have higher
codimension.

When moduli trapping acts in such a system of intersecting
enhanced symmetry loci, we expect that the moduli will first
become trapped on the locus of lowest codimension, but retain some
velocity parallel to this locus. Further trapping events can then
localize the modulus to subspaces of progressively higher
codimension.  The final result is that the moduli come to rest on
a locus of maximally enhanced symmetry.

The simplest examples of this phenomenon are toroidal
compactification, in which all circles end up at the self-dual
radius, and the system of $N$ D-branes discussed in \S2.3, in
which the gauge symmetry is enhanced to $U(N)$.\foot{A toy
model for this situation, in the case of three D-branes,
has the potential ${g^2\over 2}\bigl[\chi^2_{1} |\phi_2-\phi_3|^2 +
\chi^2_{2} |\phi_1-\phi_3|^2 + \chi^2_{3}
|\phi_1-\phi_2|^2\bigr]$, where $\phi_i$ and $\chi_i$ are six
different fields. Suppose that $\phi_2$ moves through the
point $\phi_2-\phi_3=0$. This creates $\chi_1$ particles and
traps the system at $\phi_2 =\phi_3$, where $\chi_1$ is
massless. Subsequent motion of $\phi_1$ can trap it
at the point $\phi_1=\phi_2=\phi_3$, making the
remaining fields $\chi_2$ and $\chi_3$ massless.}

Quite generally, we expect that within the accessible range  in
field space, taking into account Hubble friction and the form of
the potential, moduli trapping will select the ESPs with the
largest number of light states, which often corresponds to the
highest degree of symmetry.\foot{Moreover, as we discuss in
Appendix B, the trapping effect is far more effective at ESPs for
which the $\chi$ particles do not decay rapidly.  We therefore
expect moduli trapping to select ESPs which have relatively stable
light states.}

In some very early epoch the rolling moduli can have large
velocities, so trapping can occur even at points where the
``light'' states $\chi$ have a relatively large mass, and the
enhanced symmetry is strongly broken. However, Hubble friction
inevitably slows the motion of the moduli.  Thus, trapping at late
times is possible only at ESPs with weakly-broken symmetries and
very light particles. One could speculate about a possible relation
of this fact to the mass hierarchy problem.

Note that even though we emphasized the natural role of enhanced
symmetry in moduli trapping, in fact the only strict requirement was
the appearance of new light particles at the trapping points.
In some of the many vacua of string theory, particles may be light
not because of symmetry but because of some miraculous
cancellations. Invoking such unexplained cancellations to produce a small mass
is highly undesirable.  However, moduli trapping may ameliorate this problem, as
those rare points in moduli space where the cancellation does
happen are actually dynamical attractors.

Thus, the attractive power of symmetry and of light
particles may have implications for questions involving the
distribution of vacua in string theory
\refs{\BoussoXA,\MSS,\KKLT,\douglas}. Given the strong preference
we have seen for highly-enhanced symmetry, the distribution of all
string vacua obtained by a naive counting, weighted only by
multiplicity, may be quite different from the distribution of
vacua produced by the dynamical populating process discussed in our
paper. It is therefore very tempting to speculate that some of
the surprising properties of our world, which might seem to be due to
pure chance or miraculous cancellations, in fact may result from
dynamical evolution and natural selection.

\newsec{The Moduli Problem}

One aspect of the moduli problem is that reheating and
nucleosynthesis  can be corrupted by energy locked in oscillations
of the moduli.  The source of the problem is that the true minima
of the low-temperature effective potential applicable after
inflation do not coincide with the minima of the
Hubble-temperature effective potential which is valid during
inflation.  It follows that moduli which sit in minima of the
latter during inflation will find themselves displaced from their
true, low-temperature minima once inflation is complete. The
energy stored in this displacement, and in the resulting
oscillations about the true minimum, poses problems for
nucleosynthesis.

One way to address this problem is to permit initial displacements
of the moduli, as described above, but somehow arrange that the
oscillating moduli decay very rapidly to Standard Model particles.
Alternatively, one could fix the moduli at a scale high enough
that the Hubble temperature during inflation does not destabilize
them. This may work in string models with stabilized moduli such as
\refs{\MSS,\achar,\KKLT,\GukovCY}.

Another approach to this problem \dine\ is to posit that the
moduli  sit at an enhanced symmetry point minimum of the finite-temperature
effective potential during inflation. Then, when
inflation ends, the moduli are still guaranteed to be at an
extremum of the effective potential.  If this extremum is a
minimum then the moduli have no problematic oscillations after
inflation.  Our trapping mechanism allies nicely with this idea by
providing a preinflationary dynamical mechanism which explains the
initial condition assumed in this scenario.  That is, in parts of
the universe where $\phi$ kinetic energy dominates well before
inflation, the trapping effect can explain why the moduli find
themselves in ESP minima at the onset of inflation.

\newsec{Trapped Inflation and Acceleration of the Universe}

The main motivation of our investigation was to study the behavior
of moduli in quantum field theory and string theory. However, the
results we have obtained have more general applicability. To give
an example, in this section we will study the cosmological
implications of the trapping of a scalar field $\phi$ with a
relatively steep potential.

Consider the theory of a real scalar field $\phi$ with the
effective  potential $m^2\phi^2/2$.  In the regime $\phi < M_p$
the curvature of the effective potential is greater than $H^2$,
with $H$ the Hubble parameter, so $\phi$ falls rapidly to its
minimum, and inflation does not normally occur.

We will assume that $\phi$ gives some bosons $\chi$ a mass
$g|\phi-\phi_1|$. Let us assume that $\phi$ falls from its initial
value $\phi_0 = \phi_1(1+ \alpha) < M_p$ with vanishing initial
speed.  If we take $\alpha \ll 1$ and neglect for the moment the
expansion of the universe, then $\phi$ arrives at $\phi_1$ with
the velocity $v = \sqrt{2 \alpha}\, m\phi_1$.

As $\phi$ passes $\phi_{1}$, it creates $\chi$ particles with
number density $n_{\chi }  =    {({gv})^{3/2}/ 8\pi^3}$. After a
very short time these particles become nonrelativistic, and
further motion of $\phi$  away from $\phi_1$ requires an energy $g
| \phi - \phi_1| n_{\chi }$. In other words, the effective
potential becomes
\eqn\effpot{V(\phi) \approx {1\over 2} m^2\phi^2 + g n_{\chi }
|\phi-\phi_1|  = {1\over 2} m^2\phi^2 +g^{5/2}{v^{3/2} \over
8\pi^3} |\phi-\phi_1|  .}
For $g^{5/2}{ v^{3/2} \over 8\pi^3 m^2}>\phi_1 $, the minimum of
the effective potential is not at $\phi = 0$, but at the point
$\phi_1$, where the particle production takes place. The condition
$g^{5/2}{v^{3/2} \over 8\pi^3 m^2} >\phi_1$ implies that
\eqn\condb{m <   2^{-9/2} \pi^{-6} g^{ 5 } \alpha^{3/2} \phi_1  .}
Thus, if the mass of $\phi$ is sufficiently small, the
field will be trapped near the point $\phi_1$.

To give a
particular example, take $\phi_1 \sim M_p/2$,
$\alpha \sim 1/4$.  Then $\phi$ is trapped near $\phi_1$
if
\eqn\condc{m <   10^{-6} g^{ 5 }  M_p .}
For a very light field, such as a modulus with $m \sim 10^2$ GeV
$\sim 10^{-16} M_p$, this condition is readily satisfied unless
$g$ is very small.

Once the field is trapped, it starts oscillating around $\phi_1$
with ever-decreasing amplitude, creating new $\chi$ particles in
the regime of parametric resonance.  Eventually $\phi$ transfers a
large fraction of its energy to $\chi$ particles. One can easily
check that in this model the fall of $\phi$ to the point $\phi_1$
and the subsequent process of creation of $\chi$ particles occurs
within a time smaller than $H^{-1}$, so one can neglect expansion
of the universe at this stage. This process is therefore governed
by the theory described in \S3. In particular, we may use the
estimate \numberw\ of the total number of $\chi$ particles
produced in the process.  At the end of the particle production,
the correction to the effective
potential becomes much larger than at the beginning of the process: %
\eqn\expans{\Delta V = g |\phi-\phi_1| n_{\chi} \sim v^{3/2} g^{1/2} |\phi-\phi_1| \ ,} %

Subsequent expansion of the universe dilutes the density of $\chi$
particles as $a^{-3}$, which eventually makes the correction to
the effective potential small, so that
$\phi$ starts moving down again.
The field $\phi$ remains trapped at $\phi = \phi_1$ until the
scale factor of the universe grows by a factor
%
%
\eqn\scf{a\sim \alpha^{1/4}\left({g\phi_1\over m} \right)^{1/6}}
since the beginning of the trapping process.

In the beginning of the first e-folding, the kinetic energy of the
$\chi$ particles and of the oscillations of $\phi$ is comparable
to the potential energy of $\phi$. However, the kinetic energy
rapidly decreases, and during the remaining time the energy is
dominated by the potential energy $V(\phi_1)$. This means that the
trapping of $\phi$ may lead to a stage of inflation or
acceleration of the universe, even if the original potential
$V(\phi)$ is too steep to support inflation.

Let us consider various possibilities for the scales in the
potential,  to get some simple numerical estimates for the
duration of inflation.  For example, if we take $\alpha, g =
{\cal{O}}(1)$, $\phi_1 \sim M_p$ and $m \sim 10^2$ GeV, then the
scale factor during a single trapping event will grow by a factor
of $e^6$. If one considers a model with $m \sim 10^{-30}M_p$,
which can arise in a radiatively stable manner (as in the ``new
old inflation'' model \gs), the scale factor during a single
trapping event can grow by a factor of $e^{11}$. Finally, if the
moduli mass is of the same order as a typical mass taken in
theories of quintessence, $m \sim 10^{-60} M_p$, we can have an
accelerated expansion of the universe by a factor $e^{23}$, in a
sub-Planckian regime of field space, just from trapping. (In this
last case, as in ordinary quintessence models, tuning is
required.)

Thus, the stage of inflation in this simple model is shorter than
the usual 60 e-folds, but it may nevertheless be very useful for
initiating a first stage of inflation in theories where this would
otherwise be impossible, or for diluting unwanted relics at the
later stages of the evolution of the universe.  Moreover, this
scenario can easily describe the present stage of acceleration of
the universe.

One can also make the effect more substantial by constructing a
more complicated scenario, consisting of a chain of $N$ particle
production events at locations $\phi= \phi_i$, where some fields
$\chi_i$ become light.  The field $\phi$ may be trapped and enter
the stage of parametric resonance near each of these points.
Correspondingly, the universe enters the stage of inflation many
times.  One could arrange for 60 e-folds of inflation by taking,
for example, $m \sim 10^2$ GeV, $N \sim 10$.

\ifig\inflates{The D-brane picture of a series of trapping
events.} {\epsfxsize4.5in\epsfbox{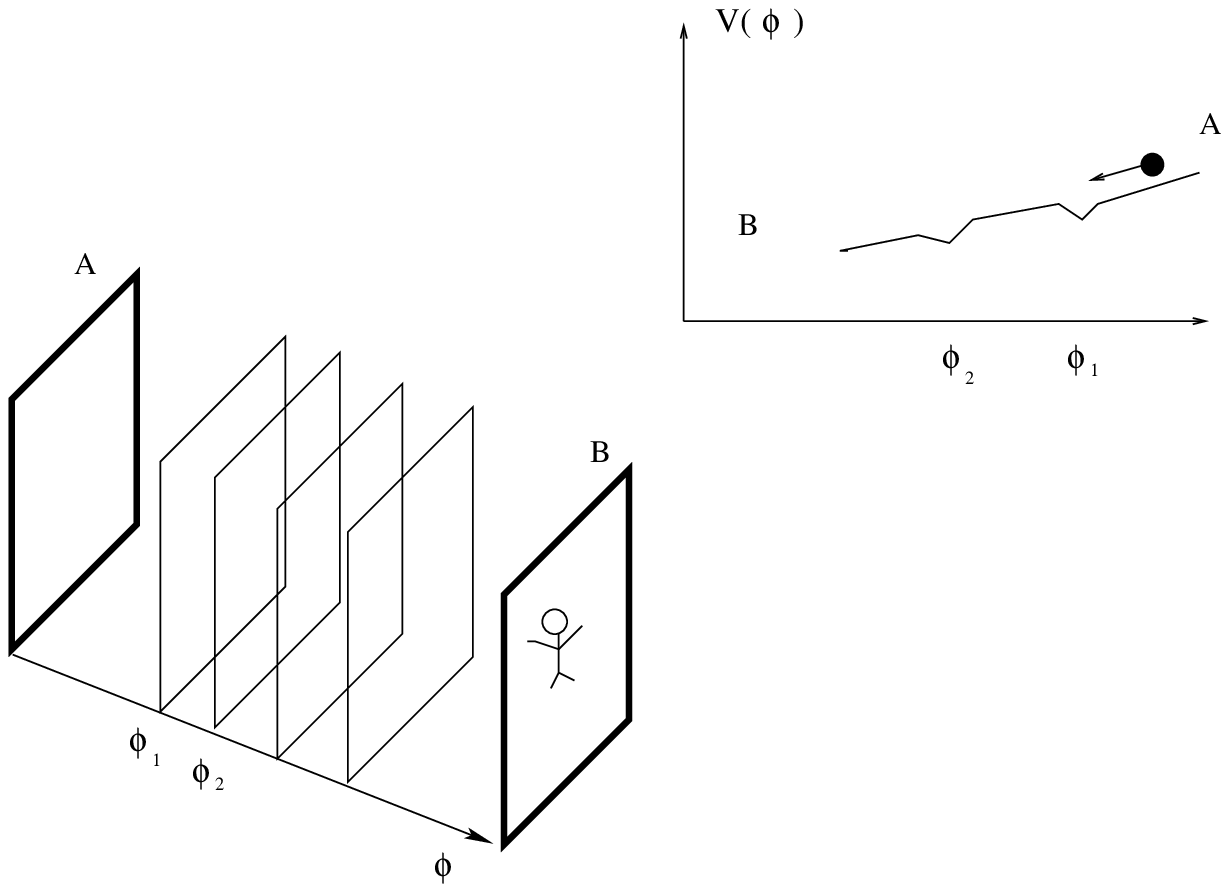}}

A D-brane example provides a useful geometrical model of this
process. Suppose we have an observable brane $B$ and another brane
$A$ approaching it. Suppose also that there are a number of other
branes in between $A$ and $B$. Each time the moving brane passes
through one of the standing intermediate branes, stretched strings
are created and slow the motion of $A$. The cumulative effect of a
number of standing branes is perceived on the observable brane as
a slowing-down of the motion of $A$ due to the interactions.

One should note that inflation in our scenario is rather unusual:
the inflaton $\phi$ rolls a short distance, then oscillates for a
long time, but with period much smaller than $H^{-1}$, then rolls
again, etc. This may lead to peculiar features in the spectrum of
density perturbations. One can avoid these features if the points
$\phi_i$ are very close to each other, and each of them does not
stop the rolling of $\phi$ but only slows it down. In this case,
particle production will not lead to parametric resonance, so it
is not very important to us whether the fields $\chi_i$ are bosons
or fermions, as long as their masses vanish at $\phi_i$.

This scenario is similar to the string-inspired thermal inflation
considered in \berera\ (see also \mavan), but our proposal does
not require thermal equilibrium. The main effect which supports
inflation in our scenario is based on particle production and has
a nonperturbative origin. (A closely-related mechanism uses  the
corrections to the kinetic terms in the strong coupling regime,
where the particle production is suppressed \SilversteinHF.) We
hope to return to a discussion of this possibility in a separate
publication.

\newsec{Conclusion}

We have argued that the dynamics of rolling moduli is considerably
modified due to quantum production of light fields.  In flat space
quantum field theory, moduli typically become trapped in orbits
around loci which have extra light degrees of freedom. In the
presence of gravity, Hubble friction limits the field range the
system samples, but any trapping events which do occur are
enhanced by Hubble friction, which rapidly brings the modulus to
rest at an ESP.  Moduli trapping may aid in solving the
cosmological moduli problem by driving moduli to sit at points of
enhanced symmetry. Furthermore, the trapping of a scalar field
which has a potential can lead to a short period of accelerated
expansion in situations with steeper potentials than would
otherwise allow this.  Finally, the trapping effect has important
consequences for the problem of vacuum selection, as it can reduce
the problem to that of selecting one point within the class of
ESPs. An intriguing feature of this process is that the trapping
is more efficient near points with a large number of unbroken
symmetries.

\medskip\

\noindent{\bf Acknowledgements}

We would like to thank M. Dine, G. Felder, S. Kachru, R. Kallosh,
R. Myers, S.-J. Rey, S. Shenker, D. Tong, and A. Tseytlin for
useful discussions. L.K. is supported by NSERC and CIAR. The
remaining authors are supported in part by the DOE under contract
DE-AC03-76SF00515 and by the NSF grant PHY-0244728.

\bigbreak\
\bigbreak\

\appendix{A}{Particle Production Due to Motion on Moduli Space}

In this section we will calculate the quantum production of
$\chi$ particles, ignoring the effect of backreaction on the motion of
$\vphi$.

A mode of $\chi$ with spatial momentum $k$ obeys the wave equation
\eqn\ueqq{ \Bigl(\p^2_t+ k^2 + g^2(\mu^2 + v^2 t^2)\Bigr) u_k =0.}
There are two solutions to this equation, $u^{in}_k$ and
$u^{out}_k$, associated to vacuum states with no particles in the
far past and no particles in the far future, respectively. These
two sets of modes are related by a Bogoliubov transformation
\eqn\asd{ u^{in}_k = \alpha_k u^{out}_k + \beta_k u^{out}_k{}^*.}
If we start in the state with no particles in the far past, then
one can calculate the number density of particles in the far
future to be \eqn\asdt{ n_k = |\beta_k|^2} in the $k^{th}$ mode.
This may be evaluated by solving equation \ueqq\ in terms of
hypergeometric functions (see e.g. \S3.5\ of \bd), but we will
present here a more physical argument.

One can view \ueqq\ as a one dimensional Schr\"odinger equation
for particle scattering/penetration through an inverted parabolic
potential. If we send in a wave $\psi_k^{in}$ from the far right
of the potential, part of it will penetrate to the far left, with
an asymptotic amplitude $T_k \psi_k^{out}$, and part of it will be
reflected back to the right, with an asymptotic amplitude $R_k
\psi_k^{in*}$, where $T_k$ and $R_k$ are the transmission and
reflection amplitudes.\foot{The modes in the two problems are
related by $u_{k}^{in}(t\to -\infty) =
T_{k}^{*}{\psi_{k}^{out}}^{*}$, etc.}

The Bogoliubov coefficient in \asd\ is determined in terms of
these transmission and reflection amplitudes via
%
\eqn\betafrac{\beta_k = {R_k^*\over{T_k^*}}.}

Now we use a trick from quantum mechanics
to relate R and T using the WKB method. If we are moving along the
real time coordinate, the WKB form of the solution $u_k^{in}(t)$
will be violated at small $t$, due to non-adiabaticity. However,
if we take $t$ to be complex then we can move from $t=-\infty$ to
$t=+\infty$ along a complex contour in such a way that the WKB
approximation \eqn\wkb{ u_k^{in}(t) \sim {1\over
\sqrt{2\sqrt{k^2+g^2(\mu^2 + v^2t^2) }}}e^{-i\int^t
\sqrt{k^2+g^2(\mu^2 + v^2{t'}^2) }dt^\prime}}
%
is valid. Here the integral $\int^t dt'$ becomes a contour
integral along a semicircle of large radius in the lower complex
$t$ plane. For large $|t|$, we can estimate the phase integral in
\wkb\ by expanding \eqn\expand{ \sqrt{k^2+g^2(\mu^2 + v^2t^2)}
\sim gvt+{k^2+g^2\mu^2\over{2gvt}}.}
As we go around half of the circle, this term generates a factor
\eqn\factoris{(e^{-i\pi})^{-i(k^2+g^2\mu^2)/2gv -1/2} = i e^{-\pi
(k^2+g^2\mu^2)/2gv}.} This is exactly the 
ratio between $R^*$ 
and $T^*$, so we find
\eqn\simplebeta{n_k = |\beta_k|^2= e^{-\pi (k^2+g^2\mu^2)/gv}. }
%
%
%
It is important to note that this result applies much more
generally  than for \eqn\phiiss{\phi=i\mu+vt.} In many cases the
nonadiabaticity is only appreciable near the origin $\phi=0$, so
that the near-origin trajectory can be approximated by \phiiss\
with some appropriate near-origin velocity $v$, even if the
evolution away from the origin is very different from \phiis.

Moreover, in analogous circumstances (T-dual in the brane
context) with a nontrivial electric field, we obtain a similar
expression due to Schwinger pair production; a related point was
made in \BachasKX. In addition, formula \simplebeta\ applies not
only to scalar fields, but also to fields of arbitrary spin. From
this universal behavior, it is tempting to speculate that
\simplebeta\ could provide an effective model for string theory
effects, but we will not pursue this direction here.

The result \simplebeta\ is nonperturbative in $g$ (with $g$, not
$g^2$, appearing in the denominator of the exponent); it is
interesting to ask whether there is a simple interpretation of
this nonanalytic, nonperturbative effect. Similarly, it is
interesting to note that as discussed in \S3, the potential for
$\phi$ induced by particle production is linear, so that if
extended to the origin it would have a nonanalytic cusp there.

Our results correspond to the low-velocity limit of the D-brane
calculation by Bachas \BachasKX. There, he obtains an imaginary
part to the action for moving D3-branes of the form
\eqn\numisnow{ {\rm{Im}}~S \propto
 \sum _{n=1}^{\infty} {(-1)^{(n+1)}\over{n}} \left({gv\over{\pi 
n}}\right)^{3/2}
   \exp (-n \pi g\mu^2/v) }
where we have translated his results into our variables. The first
term in this expansion is proportional to the overlap $\int
d^3\vec k |\beta_k|^2$ giving the number density \numdens\ of
produced particles; this agrees with what we expect from
unitarity.  More generally, backing away from this low-velocity
limit, the calculation in \BachasKX\ combined with unitarity may
provide a simple generalization of our results to the string case,
as we briefly discussed in \S5.2.

\appendix{B}{Annihilation of the $\chi$ Particles}

In this section we study the effects of collisions and direct
decays of the created $\chi$ particles, and demonstrate that for
suitably chosen parameters the trapping effect receives only small
corrections.  More specifically, we place limits on the reduction
of the $\chi$ number density through processes like $\chi\chi \to
\phi \bar{\phi}$ and $\chi \to \eta\bar{\eta}$, where $\eta$ is
some light field.

Direct decays, if present, could easily ruin the trapping
mechanism: if the $\chi$ particles decay too rapidly into light
fields then the energy stored in created $\chi$ particles will not
suffice to stop $\phi$. In this case the modulus will roll past
the ESP, feel a transient tug toward the ESP while the $\chi$
particles remain, and then gradually break free and glide off to
infinity at a reduced speed.

We will therefore consider only models in which couplings
of the  form $\chi \bar\psi\psi$, with $\bar\psi,\psi$ very light,
are negligible.  As an example, one can easily exclude such decays
in a supersymmetric model with a superpotential of the form ${\cal
W}\sim g\Phi {\rm{X}^2}$.  Here $X$ is a chiral superfield with
scalar component $\chi$ and fermion $\psi_\chi$, and $\Phi$ is a
chiral superfield with scalar component $\phi$ and fermion
$\psi_\phi$. This generates Yukawa couplings of the form $\chi
\psi_{\chi}\psi_\phi$ and $\phi\psi_\chi\psi_\chi$, which do not
allow decays from a component of ${\rm X}$ to purely $\Phi$
particles. Thus, if all components of ${\rm X}$ are heavy, the
${\rm X}$-particle energy density we produce cannot decrease by
direct decays.  In some of the simplest brane setups, exactly this
situation is realized: a string which is heavy because it
stretches between two branes separated in a purely closed string
bulk space cannot decay perturbatively into two light, unstretched
strings.

On the other hand, a priori we cannot ignore the coupling $
{g^2\over 2} \chi^2 \phi^2$
as it is this which gives rise to the desired trapping effect.
This means that we must tolerate a certain rate of annihilation
(as opposed to direct decay).  We will now review the cross
section for this process and determine its effect on the number
density $n_{{}_{\chi}}$ appearing in
\endens.

The Lorentz-invariant cross section for the annihilation process $
\chi  \chi \rightarrow \phi \bar{\phi}$ is, written in terms of
center of mass variables, \eqn\crosssection{\sigma = {g^4 k' \over
4\pi k E^2},} where $k$ and $k'$ are the momenta of the ingoing
and outgoing particles and $E$ is the energy of the ingoing $\chi$
particles.  The reverse process $\phi \bar{\phi} \rightarrow \chi
\chi $ tends to enhance the trapping effect.  As we are in search
of a lower bound on the number of $\chi$ particles, we will simply
omit this reverse process.

We now determine the annihilation rate to find the rate at which
$\chi$ particles are lost. If we assume that all the $\chi$'s are
produced at $t=0$, we find \eqn\decayrho{{\dot n
(\vec{k}_1,t)\over n(\vec{k}_1,t)} = - \int d \vec{k}_2
n(\vec{k}_2,t){\sqrt{(k_1 k_2)^2-m_{\chi}^4} \over E_1 E_2}
\sigma(\vec{k}_1,\vec{k}_2).} Here $u={\sqrt{(k_1
k_2)^2-m_{\chi}^2}/E_1 E_2}$ is the Lorentz-invariant relative
velocity of the initial $\chi$'s and $\sigma(\vec{k_1},\vec{k_2})$
is the cross section, to be calculated using \crosssection\ in the
center of mass frame.

We can simplify \decayrho\ to get an upper bound on how fast
$\chi$ decays. Ignoring the momentum dependence on the right hand
side of \decayrho, which amounts to taking the non-relativistic
limit, and ignoring the mass of $\phi$ produced by the $\chi$
particles, we have \eqn\decayrhob{{\dot n (\vec{k}_1,t)\over
n(\vec{k}_1,t)} \geq - {g^4 \over 2\pi m_\chi^2}\int d \vec{k}_2
n(\vec{k}_2,t).} We can bound the integral in the second term on
the right hand side by $n_{{}_{\chi}}$, the total number of
$\chi$'s produced, as given in \numdens. To approximate the
time-dependence of the mass $m_{\chi}$, we take $m_\chi^2 = \mu^2
+v^2 t^2$, which is what the uncorrected motion for $\phi$ would
give. So we have finally
\eqn\rhot{{\dot n (\vec{k_1},t)\over n(\vec{k_1},t)} \geq - {g^4
n_{{}_{\chi}} \over 2\pi (\mu^2+v^2 t^2)}, }
which yields
\eqn\rhott{{n(\vec{k},t) \over n(\vec{k},0)}\geq \exp\left(-{g^4
n_{{}_{\chi}} \over 2\pi \mu v} \arctan {vt \over \mu}\right).}
This is clearly bounded from below by
\eqn\basicbound{\exp \Bigl(-{g^4 n_{{}_{\chi}} \over 2\mu
v}\Bigr).}
so that the number density is reduced over time by at worst the
factor \basicbound.

The total energy density in the $\chi$ particles at a given time
is therefore

\eqn\energyw{\eqalign{E = & \int d \vec{k}n(\vec{k},t)  \sqrt{k^2
+ g^2  (\mu^2 + v^2 t^2)}
 \cr \geq & n_{{}_{\chi}} \sqrt{ g^2  (\mu^2+ v^2 t^2)}  \exp\left(-{g^4
 n_{{}_{\chi}}
\over 2\pi \mu v} \arctan {vt \over \mu}\right). }} From this we
see that the mass amplification effect of the $\chi$ particles
inevitably prevails and stops $\phi$ from rolling arbitrarily far
past the enhanced symmetry point.

This reduction of the number density softens, but does not ruin,
the trapping effect.  Using the energy density \energyw\ in the
simple estimate leading to \matchen, we find a new estimate for
$\phi_*$:
\eqn\newmatch{\phi_{*}= {{4 \pi^3}\over{g^{5/2}}}{v^{1/2}}e^{\pi g
\mu^2/v}e^{g^4 n_{{}_{\chi}}/4\mu v}}
Thus, although collisions never lead to an escape, they do lead to
a  somewhat increased stopping length $\phi_{*}$.

For suitably chosen parameters we can arrange that the effect of
collisions is unimportant and the estimates \matchen,\newmatch\
approximately agree.  For example, the final exponential factor,
which encodes the consequences of annihilations, will be less
important than the factor $e^{\pi g\mu^2/v}$ as long as $g^3 v \ll
\mu^2$.

We conclude that direct decays can be forbidden using symmetry,
whereas collisions increase the stopping length
\newmatch\ but do not ruin the trapping effect.

\appendix{C}{Bending Angle and Energy Loss during the
First Pass}

In this section we will demonstrate that one can obtain analytical
results for the early stage of the motion of the moduli through a
systematic expansion and iteration procedure. Specifically, we
start with the uncorrected trajectory for the moduli, work out the
production of $\chi$ particles to the second order in $\dot \omega
/ \omega^2$, feed this back into the equation of motion for the
modulus, and then obtain the corrected trajectory for the modulus
to leading nontrivial order of the iteration and $\dot \omega /
\omega^2$.

The solution shows deflection and energy loss, as we would expect.
However, without going to higher orders in the iteration, and
without being able to detect particle production which is
nonperturbative in $\dot \omega / \omega^2$ (as in \nkis), we do
not expect our solution to capture the full dynamics.  In
particular, the leading-order calculations presented here do not
capture the late-time behavior of the system, where any small
particle production, such as that in \nkis, can alter the
trajectory from an escaping one to a trapped one.

We choose to work with the simple model \lis\ in which the complex
modulus $\phi=\phi_1 + i \phi_2$ has an ESP at the origin, where a
real scalar field $\chi$ becomes light. Without including
backreaction, the trajectory \phiis\ leads to a time-varying
frequency for the $\chi$ excitations, \eqn\firstomega{\omega_k(t)
= \sqrt{k^2 + g^2(\mu^2 + v^2 t^2 )}.}

We start by defining the generalized Bogoliubov coefficients
$\alpha_{\vec{k}}(t)$ and $\beta_{\vec{k}}(t)$:\foot{Similar
calculations appeared in \kls.} \eqn\abone{\eqalign{
u_{\vec{k}}(t)&={1 \over
\sqrt{2\omega_{\vec{k}}(t)}}\left(\alpha_{\vec{k}}(t) \exp\left(-
i \int^t \omega_{\vec{k}}(t') dt'\right) + \beta_{\vec{k}}(t)
\exp\left(+ i \int^t \omega_{\vec{k}}(t') dt'\right)\right) \cr
\dot u_{\vec{k}}(t) &= -i \sqrt{{\omega_{\vec{k}}(t) \over
2}}\left( \alpha_{\vec{k}}(t) \exp\left(- i \int^t
\omega_{\vec{k}}(t') dt'\right) - \beta_{\vec{k}}(t) \exp\left(+ i
\int^t \omega_{\vec{k}}(t') dt' \right)\right) .}} This is
possible as long as $\omega_k(t)$ is never zero. The consistency
of this definition combined with the Klein-Gordon equation for
$u_{\vec{k}}(t)$ leads to two coupled first-order ODE's for
$\alpha_{\vec{k}}(t)$ and $\beta_{\vec{k}}(t)$ \eqn\alph{\eqalign{
\dot {\alpha_{\vec{k}}}(t) &= {\dot \omega_{\vec{k}}(t) \over 2
\omega_{\vec{k}}(t)} \exp\left(+2 i \int^t \omega_{\vec{k}}(t')
dt'\right) \beta_{\vec{k}}(t)\cr \dot {\beta_{\vec{k}}}(t) &=
{\dot \omega_{\vec{k}}(t) \over 2 \omega_{\vec{k}}(t)}
\exp\left(-2 i \int^t \omega_{\vec{k}}(t') dt'\right)
\alpha_{\vec{k}}(t).}} The normalization for $u_{\vec{k}}(t)$
implies \eqn\norm{\alpha_{\vec{k}}^2(t) - \beta_{\vec{k}}^2 (t)
=1.} Finally, the initial condition for $\chi(t)$ as $t\rightarrow
-\infty$ corresponds to \eqn\initial{\alpha_{\vec{k}}=1,
\,\,\,\,\,\,\,\, \beta_{\vec{k}}=0.} When the frequency
$\omega_k(t)$ changes adiabatically, $|\dot \omega / \omega^2| \ll
1$, the exponentials approximate the positive and negative
frequency modes of the system and $\beta_k(t)$ describes particle
production.

In terms of $\alpha_{\vec{k}}(t)$ and $\beta_{\vec{k}}(t)$
\eqn\wsquare{\la \chi^2(t)\ra={1 \over 4 \pi^2}
\int_0^{\infty}{{dk k^2} \over \omega_k(t)}\left(1 + 2
|\beta_k(t)|^2 + 2 {\rm{Re}} \left(\alpha_k(t)
\beta_{k}^*(t)\exp\left(-2i\int^t
\omega_k(t')dt'\right)\right)\right). } The first term in the
above equation is the ordinary Coleman-Weinberg potential (after
cancelling against some polynomial counterterms), and we ignore it
for the reasons mentioned in \S3.2. The second and third terms
encode the particle production effects; we will calculate these by
an adiabatic approximation.

To the leading nontrivial order, $\alpha_k(t)$ and $\beta_k(t)$
satisfy \eqn\onealpha{\eqalign{ \dot {\alpha_{k}}(t) &= 0 \cr
\dot{\beta_{k}}(t) &= {\dot \omega_{k}(t) \over 2 \omega_{k}(t)}
\exp\left(-2 i \int^t \omega_{k}(t') dt'\right)}}
%
where we have used the zeroth-order Bogoliubov coefficients,
$\alpha^0_{\vec{k}}(t)=1,~~\beta^0_{\vec{k}}(t)=0$.
These are subject to the initial conditions \initial, so that
$\alpha_k(t)=1$ and \eqn\solbeta{\beta_k(t)= \left({i \dot
\omega_k \over 4 \omega_k^2} - {{\dot \omega_k}^2 \over 4
\omega_k^4} + {\ddot \omega_k \over 8 \omega_k^3} +... \right)
\exp(-2i\int^t \omega_k dt').} This yields \eqn\wsqu{\eqalign{\la
\chi^2(t) \ra - \delta_M & ={1 \over 16 \pi^2} \int_0^{\infty} dk
k^2 {1 \over \omega_k(t) }\left({\ddot \omega_k(t) \over
\omega_k(t)^3} -{3 {\dot \omega_k(t)^2} \over 2 \omega_k(t)^4}+...
\right) \cr &= { v^2 \mu^2 \over 48 \pi^2 (\mu^2+ v^2 t^2)^2}.}}
To the first order of iteration, the corrected equation of motion
for $\phi$ is

\eqn\psieom{\ddot \phi + {g^2 v^2 \mu^2 \over 48 \pi^2  (\mu^2
+v^2 t^2 )^2}\phi=0} with initial conditions\eqn\init{\phi(t_0)=v
t_0 + i \mu,\,\,\,\,\,\,\,\,\,\, \dot \phi(t_0)=v.}  We will
eventually take the initial time $t_0$ to be $-\infty $. Moreover,
we have in mind $v_0<0$, $\mu>0$, so that $\phi$ comes from
$\phi_1 = + \infty$ with a displacement in the positive direction
of $\phi_2$ and shoots towards the left.

The solution for the above system is \eqn\solu{\phi(t) =  \mu e^{
{i \pi \sqrt{1+c}} \over 2} \left((1 + {1 \over \sqrt{1+c}})
\varphi_1({\mu \over vt}) +  (1 - {1 \over \sqrt{1+c}})
\varphi_2({\mu \over vt})\right),} where \eqn\para{\eqalign{ c &=
{g^2 \over 48\pi^2}\cr \varphi_1(z) &= {1 \over 2z} (z-i)^{{1 +
\sqrt{1+c} \over 2}}(z+i)^{1 - \sqrt{1+c} \over 2},\cr
\varphi_2(z) &= {(-1)^{-\sqrt{1+c}} \over 2z} (z-i)^{{1 -
\sqrt{1+c} \over 2}}(z+i)^{1 + \sqrt{1+c} \over 2},}} and the
branch cut goes from $z=-i$ to $z=i$.

As time goes from $-\infty$ to $+\infty$, $z$ goes from $0^+$  to
$\infty$ and then to $0^-$. We can enclose the contour by a
semicircle either in the upper or lower half-plane, giving the
same result: \eqn\monodroy{\eqalign{\varphi_1 & \rightarrow e^{i
\pi (1 + \sqrt{1+c})}\varphi_1 \cr \varphi_2 & \rightarrow e^{i
\pi (1 - \sqrt{1+c})}\varphi_2 .}} We find that at late time
\eqn\asymp{\eqalign{\phi(t \rightarrow +\infty) \rightarrow & -v t
\Bigl(\cos(\pi \sqrt{1+c}) + {i \over \sqrt{1+c}} \sin(\pi
\sqrt{1+c})\Bigr) \cr & - i \mu \Bigl(\cos(\pi \sqrt{1+c}) + i
\sqrt{1+c} \sin(\pi \sqrt{1+c})\Bigr) \cr & + {\cal{O}}({\mu \over
vt }).}} The leading large time behavior is captured by the first
term, which describes a constant linear motion on the moduli
space. It shows a deflection in the direction of the ESP
(counterclockwise) determined by the coefficient of the leading
term $\Bigl(\cos(\pi \sqrt{1+c}) + {i \over \sqrt{1+c}} \sin(\pi
\sqrt{1+c})\Bigr)$. As we expect, if $g=0$ the modulus goes
through a straight line from right to left.  The bending angle
increases monotonically, with no upper bound, as $g$ increases.
Thus, if $g$ is very big, $\phi$ can orbit around the origin many
times, with a growing radius, until eventually it shoots out to
infinity in a fixed direction.

The modulus loses a finite fraction of its energy as it passes by
the ESP. This can be seen from the fact that the coefficient of
$-vt$ has norm less than one in \asymp. It so happens that at this
order of iteration the fraction of energy lost depends only on the
coupling $g$: \eqn\energyloss{-\Delta E /E_0  = {c \over 1+c }
\sin^2(\pi \sqrt{1+c}).}

The above picture is valid only in certain parameter regimes.
Specifically, if taking only the leading nontrivial contribution
to $\la \chi^2(t) \ra$ is to make any sense, then the
non-adiabaticity parameter must be small:

\eqn\nonadi{\left|\dot \omega_k \over \omega_k^2 \right|\sim {v
\over g \mu^2}\ll 1.} This condition also implies that the effects
of the produced $\chi$ particles on the motion of $\phi$ shows up
only at exponentially late times, so that it is sensible to have
an intermediate stage in which the $\phi$ field has a linear
motion. Of course, we know this is not the full picture. The
linear growth of the mass of the $\chi$ particles would eventually
stop $\phi$ and bring it back.

\appendix{D}{Classical Trapping Versus Quantum Trapping}

In this section we will compare our quantum trapping mechanism
with the purely classical trapping proposed in \HellingKZ.

Consider for simplicity a theory of two real scalar fields, $\phi$ and
$\chi$, with the interaction ${g^2\over 2} \phi^2\chi^2$.  In our
discussion in the main text we assumed the initial conditions $\la \chi
\ra=0, \la \phi \ra \neq 0, \dot{\chi}=0,\dot{\phi}=v$.  Potentially
interesting classical dynamics arises in the more general case in which
the initial velocity of $\chi$ is nonzero \HellingKZ.


Let us therefore consider the classical behavior of these fields, ignoring
particle production entirely.  If we define $v \equiv \sqrt{\dot\chi^2 +
\dot\phi^2}$, then energy conservation implies that the trajectory of
$\phi$ and $\chi$ is bounded by the surface $g^2\phi^2\chi^2 = v^2$.  The
fields will evidently start bouncing off the curved walls of the
potential.  This bouncing will be highly random.

Naively, one would expect that on average the fields become
confined in the region
\eqn\clampl{\langle\phi^2 \rangle = \langle \chi ^2\rangle \sim
{v\over g}.}
This result would coincide with our estimate for the amplitude of
the oscillations of $\phi$ at the end of the stage of parametric 
resonance, cf. \finalampl.

However, the situation is more complicated. As we are going to
show, the fields spend most of the time not at $|\phi| \sim |\chi|
\sim \sqrt{v\over g}$, but exponentially far away from this
region, moving along one of the flat directions of the potential.

To see this, note that because of the chaotic nature of the bouncing, it
will occasionally happen that the fields enter the valley $\chi \ll \phi$
at a small angle to the flat direction, i.e. with velocities obeying
$|\dot\chi| \ll \dot\phi$.  Defining $|\dot\chi| \equiv \alpha v$, we are
interested in the case that the angle $\alpha$ happens to be small.

Energy conservation implies that the amplitude of the oscillations of
$\chi $ at the initial stage of this process is approximately ${\alpha
v\over g\phi}$. Because of the interaction term ${g^2\over
2}\phi^2\chi^2$, these oscillations act on the field $\phi$ with an
average returning force $\sim {\alpha^2 v^2\over \phi}$, which corresponds
to the logarithmic potential $V(\phi) \sim \alpha^2 v^2 \log \phi$
\HellingKZ. Clearly, this potential will eventually pull the field
$\phi$ back to the ESP $\phi = 0$. However, this happens at
exponentially large $\phi$: the field starts moving back only after its
value approaches 
\eqn\maxclass{\phi^*_{\rm class} \sim \sqrt{v\over g}~
e^{c/\alpha},} 
where $c = {\cal{O}}(1)$.  

Once again, because the bouncing process is highly random, we do not
expect that the probability to enter the valley at a small angle $\alpha$
is exponentially suppressed.  This means that after bouncing back and
forth near the point $\phi=\chi = 0$, the fields $\phi$ and $\chi$
eventually enter one of the valleys at a small angle, and subsequently
spend a very long time there.  In general, the fields will spend an
exponentially long time at an exponentially large distance from the
origin. Thus, the classical trapping mechanism, unlike the
particle production mechanism described in our paper, does not lead to a
permanent trapping of the fields in the vicinity of the point $\phi = \chi
= 0$.

\listrefs\
\end